\documentclass[]{aa}  
\usepackage{graphicx}
\usepackage{xcolor}
\usepackage{txfonts}
\usepackage{natbib}
\usepackage{hyperref}
\usepackage{linenoaa}
\usepackage{makecell}
\usepackage{amssymb}
\linenumbers
\hypersetup{
    colorlinks=true,
    citecolor=blue,
    linkcolor=blue,      
    urlcolor=blue,
}

\makeatletter
\renewcommand*\aa@pageof{, page \thepage{} of \pageref*{LastPage}}
\makeatother

\begin{document}

\title{Volcanic satellites and ion escape in the magnetospheres \\ 
of ultra-cool and brown dwarf stars}

\author{A.~F.~Lanza
\inst{1}
\and
A.~V.~Oza\inst{2,3}
}

\institute{INAF-Osservatorio Astrofisico di Catania, Via S.~Sofia, 78 - I-95123 Catania, Italy\\
\email{antonino.lanza@inaf.it}
\and
California Institute of Technology, 1200 E California Boulevard, Pasadena, CA 91125, USA \\
\email{oza@caltech.edu}
\and
Jet Propulsion Laboratory, Pasadena, CA USA
}

\titlerunning{Ion sources in UCD stars}

\abstract{Radio emissions at  gigahertz frequencies of ultra-cool dwarf and brown dwarf stars suggest the presence of radiation belts not unlike Jupiter's. We investigate the possibility that the inferred magnetospheric plasma at the primary star is  sourced by an active planet or via ion escape, which we modeled as a weak ionospheric outflow. We consider indirect methods to estimate the magnetospheric plasma mass flow from auroral radio emission and apply them to the ultra-cool dwarf LSR~J1835+3259. We find that an ionospheric outflow on the order of $10^{5}$~kg~s$^{-1}$ is a viable source if the ionospheric effective Pedersen conductance is lower than $\sim 0.003$~mho. On the other hand, a tidally heated volcanic satellite with the same mass and radius as Io and an orbit with a semimajor axis smaller than $\sim 10$ stellar radii, whose eccentricity ($e \sim 10^{-3}$) is maintained by perturbations by other planets in the system, is found to be a viable source of magnetospheric plasma without strong limitations on the ionospheric conductance. The volcanic satellite scenario is also naturally in sync with the recent finding that ultra-cool dwarfs with distant substellar or stellar companions are remarkably more likely to be detected as radio emitters.}

\keywords{stars: magnetic fields -- brown dwarfs -- planets and satellites: aurorae -- planet-star interactions -- stars: individual: LSR~J1835+3259}

\maketitle

\nolinenumbers %uncomment to remove line numbering in this document
\section{Introduction}
\label{intro}

Ultra-cool dwarf (UCD) stars at the lower end of the main sequence \citep[e.g.,][]{Route17,KaoShkolnik24} and brown dwarfs (BDs) with rotation periods of a few hours have been detected as radio emitters \citep[e.g.,][]{Kaoetal16,Kaoetal18,Kaoetal19}. They show both a bursty radio emission and a quiescent emission when observed at frequencies in the gigahertz range. The former is usually strongly circularly polarized and is interpreted as electron-cyclotron maser (ECM) emission peaking at a frequency 
\begin{equation}
\nu_{\rm ECM} = 2.8\, B,
\label{ecm_nu}
\end{equation}
where $\nu_{\rm ECM}$ is the frequency in megahertz and $B$ the magnetic field in gauss \citep[e.g.,][]{Noyolaetal14}. Therefore, a bursty emission at 8.4 GHz implies a magnetic field of $\sim 3$~kG in the emitting region; this is likely the magnetic field at the poles close to the surface of the star, assuming the stellar field has a dipolar geometry. ECM emission is not isotropic but is directed inside a small solid angle around the surface of a hollow cone, the axis of which is along the local magnetic field and has an aperture of $\sim 70^{\circ}-80^{\circ}$ {\citep[e.g.,][]{MelroseDulk82,Zarka98}}. Therefore, ECM emission can be observed only during a short interval along each stellar rotation; it produces nearly periodic bursts with typical durations of minutes or tens of minutes as the star rotates. In addition to these slow bursts, short (milliseconds) bursts are intrinsically produced by the ECM, likely due to the rapid growth of the instability itself, as in the case of Jupiter \citep{Hessetal07} and possibly M dwarfs \citep{Zhangetal23}.  

The detection of ECM emission from UCD stars  is not limited to frequencies in the gigahertz domain. As witnessed by, for example,  \citet{Vedanthametal20,Vedanthametal23}, observations at 144 MHz have revealed ECM emission in a T-dwarf binary that implies an ion source of $\dot{M} \sim$ 2.5 $\times$~10$^{4}$ kg~s$^{-1}$  provided the B field is at least $\sim$ 50 G, with a likely magnetic field strength of several hundred gauss. This is remarkably stronger than the theoretically estimated $\sim$ 13-85~gauss magnetic field strengths in the cases of WASP-49~b, WASP-69~b, HAT-P-12~b, and HD~189733~b  \citep[e.g.,][]{Narang_2023A,Narangetal23}. 
On the other hand,  quiescent emission in UCD stars is observable continuously and persists for several years, with slow variations and a low degree of circular polarization. It is interpreted as  high-harmonic synchrotron radiation  originating at several stellar radii from the surface of the star, where the magnetic field is on the order of a few gauss. The fraction of radio-emitting UCDs and BDs is approximately 20\% after correcting for the known selection effects that affect their statistics in a volume-limited sample  \citep{KaoShkolnik24}.

A particularly interesting system, which we adopted as a benchmark in this study, is  \object{LSR J1835+3259}. Its quiescent emission has been spatially resolved at 4.94~GHz \citep{Climentetal23} and  8.4~GHz \citep{Kaoetal23}. This emission shows  two approximately symmetric lobes centered at a distance of $\sim 9$ stellar radii from the center of the star. This is indicative of radiation belts in a stellar magnetosphere governed  by a large-scale dipolar magnetic field akin to the case of Jupiter \citep[see][]{dePater81,dePaterDunn03}. 

An important open question is the source of the electrons responsible for the ECM emission and those producing the synchrotron radiation.  A phenomenon analogous to the radio emissions of the UCDs and BDs considered above is represented by the ECM and synchrotron emissions observed in the magnetosphere of \object{Jupiter}. The main source of ions in that case is the volcanic moon \object{Io}. 
It is worth noting that the first signature of volcanism on Io was detection in the radio decametric  emission \citep{Bigg64}; however, it was not seen in visible light (e.g., atomic sodium) until 1974 \citep{BrownChaffee74} and was finally confirmed by the observation of plumes by Voyager 1 \citep{Morabitoetal79}. The production of plasma on Io mainly results from electron-impact ionization, i.e., the thermal electrons of its plasma torus colliding with Io's atmospheric molecules, essentially sulfur dioxide  \citep{Sauretal99,BagenalDols23}, a component of which is also atmospherically sputtered \citep{Haffetal81, McGrathJohnson87, Johnson90} (cf. Sect.~\ref{tidal_heating_mass_loss}).  \citet{Dolsetal24} propose that  ionization by high-energy field-aligned electron beams also plays a  role, while photoionization by solar ultraviolet radiation has a secondary role, contributing less than 10-15\% of the total ionization rate \citep[][]{Sauretal99}. Moreover, charge exchange contributes to the ion--neutral balance \citep{McGrathJohnson89, BagenalDols23}. 

The mechanisms that power the auroral emission in UCD stars have been reviewed by \citet{Sauretal21}, who considered ionospheric--magnetospheric coupling (see Sect.~\ref{ionospheric-magnetospheric}) and the Alfv\'en waves excited by the motion of a planet orbiting inside the magnetosphere of the star (see Sect.~\ref{density_from_ECM}). 
Such a star--planet interaction mechanism was criticized by \citet{Kuznetsovetal12}, but its plausibility was later reassessed by  \citet{Letoetal17}. Therefore, we also considered it to estimate the mass flux rate across the magnetosphere of UCD stars. 

We limited ourselves to an analysis of the ECM emission because the use of  quiescent radio emission to make inferences on  the magnetospheric mass flux is hampered by an incomplete knowledge of the acceleration and transport mechanisms of the relativistic electrons across a stellar magnetosphere. Those relativistic electrons, responsible for the quiescent synchrotron emission, represent a tiny fraction in comparison with the thermal electron distribution, which depends critically on the details of such physical processes. Even in the case of the magnetosphere of Jupiter, our understanding of the relativistic electron population is too limited to link the synchrotron emission to the total electron (and plasma) density inside the magnetosphere \citep[cf. Sect.~4.3 of][]{Kollmanetal18}. Therefore, even if the density of radiating relativistic electrons can be estimated from the maximum flux of the quiescent emission and its corresponding  frequency \citep[e.g.,][Sect.~3.1]{Bergeretal08a}, we cannot use such information to estimate the mass flux across the magnetosphere of LSR~J1835+3259 \citep[see][for a discussion and the modeling of the quiescent emission of our star]{Metodievaetal17}. 

Further information on the magnetospheric environment of UCD stars is provided by their X-ray emission. Their X-ray luminosities  are much lower than what would be expected on the basis of the relationship between the radio and the X-ray fluxes observed in stars with a  higher mass and effective temperature, the so-called G\"udel-Benz law \citep{GuedelBenz93,Gudel04}. This is likely related to the almost neutral atmospheres of these very low-mass stars that do not allow for stellar coronae similar to those observed in higher-mass stars. Transient energy release associated with stellar flares is  rare in the stars in our study and does not appear to be a significant source of electrons  \citep[and therefore cannot account for their radio emission;][]{Kaoetal23}. On the other hand, we  investigated the possibility that a weak stellar ionospheric outflow can contribute as a plasma source to the magnetosphere (see Sect.~\ref{stellar_wind_main} and Appendix~\ref{stellar_wind}) and found it plausible under certain conditions. In addition,  we explored the role of small planets similar to Io, that is,  planets orbiting radio-emitting UCDs and BDs that can be sources of electrons. 

It is also interesting to note that active satellites orbiting hot Jupiters have been invoked as a source of metals -- in particular sodium, a neutral alkali metal with a rapid photoionization time, in the observed exospheres of transiting planets \citep{Ozaetal19}. One candidate is especially compelling based on a Doppler redshift approaching the satellite orbital velocity near $\sim$ +15 km~s$^{-1}$ \citep{Ozaetal24}. Although exoplanet atmospheres can have solar abundance alkali metals in their escaping envelopes, \citet{Schmidt22} points out that the unique vector of radiation pressure along our line-of-sight would Doppler blueshift alkali metals like at Mercury. Therefore, a Doppler redshift is unambiguous evidence of a satellite--magnetosphere interaction, where redshifts \citep{Unni2025} can be fueled by charge exchange or dissociative recombination as is the case at Io \citep{Schmidtetal23}. 
The presence of a SO$_2$ plasma torus somewhat analogous to what we describe here has been suggested around WASP-39~b on the basis of variable JWST/MIRI data \citep{Oza2026}. Furthermore, \citet{SucerquiaCuello25f} demonstrate that  close-in satellites orbiting giant planets can be stable on gigayear timescales. 

\section{Model}
\label{model}
In this section, we consider the two mechanisms that have been proposed to produce the auroral ECM emission in LSR~J1835+3259: the ionospheric--magnetospheric coupling (Sect.~\ref{ionospheric-magnetospheric}) and the star--planet interaction that excites Alfv\'en waves through the orbital motion of a planet inside the stellar magnetosphere (Sect.~\ref{density_from_ECM}). Starting from the observed ECM flux, these two models can be used to estimate an upper limit  for the mass flow, $\dot{M}$, across the stellar magnetosphere due to magnetospheric convection \citep{Hilletal81}.  In the case of Jupiter, the two  mechanisms are simultaneously in operation, with an average radio power of $\sim 10^{10}$~W for the auroral spot produced by Io and $\sim 4\times 10^{10}$~W for the overall Jovian auroral emission mainly powered by the magnetospheric--ionospheric coupling \citep{Zarkaetal01}. Nevertheless, being interested in estimating an upper limit on $\dot{M}$, we considered each mechanism as operating alone and being responsible for the entire ECM emission of LSR~J1835+3259. We refer the reader to \citet{Kuznetsovetal12} and \citet{Letoetal17} for the possibility of distinguishing between the two mechanisms on the basis of the time modulation of the observed radio flux. The magnetospheric mass flow can be supplied by a stellar ionospheric outflow and/or by a volcanically active planet. We introduce and theoretically discuss these two sources in Sect.~\ref{stellar_wind_main}, providing details in Appendix~\ref{stellar_wind}  and Sect.~\ref{planet_mass_loss_mechanisms}, respectively. Specifically, we  consider the case of a tidally heated planet, akin to the Jupiter moon Io, in Sects.~\ref{tidal_heating_mass_loss} and~\ref{ecc_maintaining}, while the case of a planet heated by electromagnetic induction is discussed in Sect.~\ref{induction_model_0} and in Appendixes~\ref{induction_model} and~\ref{electric_conduc}.

\subsection{Ionospheric--magnetospheric coupling}
\label{ionospheric-magnetospheric}

Ionospheric--magnetospheric coupling has been invoked to explain the auroral radio emission due to the ECM instability in giant exoplanets and UCD stars by \citet{Nichols11} and \citet{Turnpenneyetal17} using a model analogous to that of the oval auroral emission in Jupiter. This model is based on the flow of electric currents that couple the magnetosphere with the ionosphere of the star, powered by the deviation from rigid rotation in the magnetosphere at large distances from the star itself. Such a deviation is produced by the angular momentum flux associated with a mass flux $\dot{M}$ across the magnetosphere that in turn is due to magnetospheric convection. 

The rapid rotation of our UCD star makes the centrifugal force stronger than the gravity beyond the corotation radius $R_{\rm cor}$ that is given by 
\begin{equation}
    R_{\rm cor} = \left( \frac{GM_{\rm s}}{\Omega_{\rm s}^{2}}\right)^{1/3},
    \label{rcorot}
\end{equation}
where $G$ is the gravitational constant, $M_{\rm s}$  the mass of the star and $\Omega_{\rm s}$ its angular velocity of rotation. 
In the domain beyond the corotation radius, i.e., for $r \ge R_{\rm cor}$,  magnetospheric convection occurs, that is, magnetic flux tubes with a plasma density greater than their neighbor tubes move  outward in the centrifugal potential, while less dense tubes move inward \citep[see Sect. 5.1 of][]{BagenalDols20}. The collective effects of such interchange motions were described by \citet{SiscoeSummers81} as a diffusive process considering only interchanges occurring on small spatial  scales. However, \citet{Hilletal81} pointed out that the motion can develop itself on  length scales comparable with the size of the magnetosphere, if there is a longitudinal asymmetry in the mass distribution as it happens in the case of Jupiter due to the asymmetry of the density inside the plasma torus fed by its satellite Io. In such a case, a magnetospheric convective motion occurs with the denser longitude sectors moving outward in the centrifugal potential, while a return flow toward the planet sets  in the less dense longitude ranges. 

The deviation from rigid rotation produces  an azimuthal component of the magnetic field that is associated with an electric current that flows along the magnetic field lines, thus coupling the magnetosphere with the Pedersen layer of the stellar atmosphere (where the gyrofrequency approaches the collision frequency), that is, the ionosphere of the star, where the electric circuit is closed.  The precipitation of electrons onto the ionosphere, accelerated by the difference of potential that maintains the field-aligned electric current, is responsible for the ECM instability and the radio emission  from the high-latitude regions of the star akin to the radio emission from the Jupiter auroral ovals. 

The radial length scale beyond which the angular velocity of rotation of the magnetospheric plasma deviates from rigid corotation with the star is known as the Hill radius, $R_{\rm H}$,  and is given by
\begin{equation}
    R_{\rm H} = R_{\rm s} \left( \frac{2\pi \, \Sigma_{\rm p}^{*} \,B_{\rm s}^{2} \, R_{\rm s}^{2}}{\dot{M}}\right)^{1/4},
    \label{r_hill}
\end{equation}
where $\Sigma_{\rm p}^{*}$ is the effective Pedersen conductance in the ionosphere of the star, $R_{\rm s}$ its radius, $B_{\rm s}$ its surface equatorial magnetic field, and $\dot{M}$ the mass flux across the magnetosphere produced by magnetospheric convection \citep{Hill01, Turnpenneyetal17}. 

A full model of the ionospheric--magnetospheric currents and the process by which they power the ECM emission of the star is provided by 
\citet{Turnpenneyetal17}. Here we limited ourselves to a qualitative description of their model which allowed us to estimate the mass flux, $\dot{M}$, across the magnetosphere  required to produce a deviation from corotation that accounts for the observed ECM power. 

The ionospheric--magnetospheric coupling model depends on several parameters \citep[see Sect.~2 of][]{Turnpenneyetal17}. The angular velocity of rotation and the radius of our UCD star can be measured or estimated from its structure model, respectively.
The magnetic field can be evaluated from the frequency of the ECM emission using Eq.~\eqref{ecm_nu} and generally refers to the poles of the star where the field is the strongest. To evaluate the stellar field inside the magnetosphere, $B_{\rm m}$, we simply scaled it with the distance assuming a dipole geometry that is supported by the observations of the resolved magnetosphere around LSR J1835+3259 \citep{Climentetal23,Kaoetal23}. 

On the other hand, the effective Pedersen conductance, $\Sigma^{*}_{\rm P}$, the electron number density, $n$, and the plasma particle thermal energy, $W_{\rm th}$, are poorly known. For this reason, \citet{Turnpenneyetal17} explore a range of variation for those parameters considering 
$0.01 \leq \Sigma^{*}_{\rm P} \leq 10$~mho, $10^{3} \leq n \leq 10^{5}$~m$^{-3}$, and $0.25 \leq W_{\rm th} \leq 25$~keV.  
The wide range adopted for the Pedersen conductance, equivalently 0.01-10 Siemens, is justified because the electron precipitation associated with the electric currents propagating along the magnetic field lines can enhance the conductivity in a planetary or stellar atmosphere. Specifically, in the case of Jupiter, \citet{NicholsCowley04} found an enhancement from $\sim 0.05$ to $0.7$~mho. A similar enhancement can occur in the case of UCD stars, although a quantitative determination of the effective Pedersen conductance in their atmospheres is still lacking \citep[cf. Sect.~4 of][]{Turnpenneyetal17}. 

Concerning the other parameters $n$ and $W_{\rm th}$,  the values measured outside the current sheet of Jupiter are $n= 10^{4}$~m$^{-3}$ and $W_{\rm th} = 2.5$~keV \citep{Scudderetal81}, in the middle of the ranges we explored.
Moreover, the stellar magnetosphere can be assumed to be closed as in the case of Jupiter or its magnetic field lines can be assumed to be open above some latitude and interacting with the interstellar medium as the star moves across the medium itself. These two different magnetospheric configurations imply different mass flow rates, $\dot{M}$, for a given ECM emission flux (cf. Sect.~\ref{mdot_ionospheric-magnetospheric}).

\subsection{Star--planet interaction}
\label{density_from_ECM}
The  ECM emission that is observed as slow auroral bursts likely  originates  close to the poles of the star where the field is strongest \citep{Kaoetal23}. In the framework of the star--planet interaction model \citep{Sauretal21}, ECM is powered by Alfv\'en waves excited by the orbital motion of a close-by planet. The energy made available by those waves  can be estimated by means of the Alfv\'en wing model \citep[][]{Neubauer80}. 

We assumed that the round-trip time of an Alfv\'en wave from the planet to the stellar surface is longer than the convection time of the plasma past the planet. This is the case in the relatively dense magnetospheres that we considered to estimate the maximal mass flow. The opposite regime (the so-called unipolar inductor regime) will not be considered here and we refer to \citet{Neubauer98} and \citet{Sauretal21} for its treatment. 

The expression in Eq.~(8) of \citet{Noyolaetal14}, discussed in \citet{Narangetal23}, provides the radio flux $S$ observed at the Earth in the case of star--planet interaction as
\begin{equation}
S = \frac{2\pi \, \beta_{\rm S} \, R_{\rm m}^{2}B_{\rm m}^{2}V_{0}}{\mu \, \nu_{\rm ECM} \, \Omega \, d^{2}} \sqrt{\frac{\rho_{\rm m}}{\rho_{\rm m} + \mu^{-1} (B_{\rm m}/V_{0})^2}},
\label{secm}
\end{equation}
where $\beta_{\rm S}$ is the conversion efficiency between the Alfv\'en wing power and the radio power, $R_{\rm m}$ the radius of the planet, $B_{\rm m}$ the stellar magnetic field strength at the planet, $V_{0}$ the relative velocity between the local magnetic field and the planet along its orbit, $\rho_{\rm m}$ the plasma density at the planet, $\mu$ the magnetic permeability of the vacuum, $\nu_{\rm ECM}$ the ECM frequency (see Eq.~\ref{ecm_nu}), $\Omega$ the solid angle within which the emission is beamed, and $d$ the distance of the star from the Earth. {The conversion efficiency $\beta_{\rm S}$ can be assumed equal to that of the Jupiter--Io system, that is, on the order of 0.01 \citep{Zarkaetal01,Zarka07}; similarly, the solid angle of the beamed ECM emission is assumed to be $\Omega =0.15$~sr \citep{QueinnecZarka01,Zarkaetal04}.} The relative velocity $V_{0}$ in the case of a circular prograde planetary orbit in the equatorial plane of the star is $V_{0}=| \Omega_{\rm s} a - \sqrt{GM_{\rm s}/a}|$, where $\Omega_{\rm s}$ is the stellar spin angular velocity, $G$ the gravitational constant, $M_{\rm s}$ the mass of the star, and $a$ the orbital radius. 

 When applied to the case of the Jupiter--Io system, Eq.~\eqref{secm} implies an emitted radio power of $\sim 5 \times 10^{9}$~W adopting $\beta_{\rm S}=0.01$, $V_{0} \sim 57$~km~s$^{-1}$, $B_{\rm m} = 0.02$~Gauss \citep{Zarkaetal01} and a density of the plasma around Io of $\rho_{\rm m} \sim 7 \times 10^{-17}$~kg~m$^{-3}$ \citep[see Sect.~2.4 of][]{Noyolaetal14}. Such an estimate is of the same order of magnitude of the average observed radio power as produced by the interaction of Io with Jupiter's magnetic field \citep[cf. Sect.~2 of][]{Zarkaetal01}.

Equation~(\ref{secm}) allows us to estimate the density of the plasma, $\rho_{\rm m}$, close to the planet  provided that the flux of the auroral radio emission is measured, the radius of the planet, $R_{\rm m}$, is assumed, and the stellar magnetic field at the distance of the planet, $B_{\rm m}$, is known, typically by adopting a dipole geometry \citep[cf. for example,][]{Narangetal23}. 
The sensitivity to the density $\rho_{\rm m}$ of Eq.~\eqref{secm} is maximal when $\rho_{\rm m}$ is much smaller than the critical density $\rho_{\rm crit } \equiv  \mu^{-1} (B_{\rm m}/V_{0})^{2}$, a regime for which $S \propto \sqrt{\rho_{\rm m}}$ as discussed by \citet{Noyolaetal14}. 

It is important to note that the value of $B_{\rm m}$ decreases with the distance from the star making the density $\rho_{\rm m}$ diverge at the orbital distance where
\begin{equation}
    B_{\rm m} = B_{\rm lim} \equiv \sqrt{\frac{\mu\, S \,\nu_{\rm ECM} \, \Omega \, d^{2}}{2\pi \beta_{\rm S} \, R_{\rm m}^{2} \, V_{0}}}.
    \label{dist_div}
\end{equation}
Therefore, Eq.~\eqref{secm} cannot be applied beyond that limit for our purposes. Moreover, another divergence occurs at the corotation distance $R_{\rm cor}$ where $V_{0}=0$. Nevertheless, in the case of fast-rotating UCD stars, $R_{\rm cor}$ is usually inside the Roche limit (within which a planet is destroyed by the tidal force of the star), which is given by \citep[cf.][]{FordRasio06}
\begin{equation}
    a_{\rm Roche} = 2.16 \, R_{\rm m} \left( \frac{M_{\rm s}}{M_{\rm m}}\right)^{1/3},
\end{equation}
where $M_{\rm m}$ is the mass of the planet. 

When $\rho_{\rm m} \ga \rho_{\rm crit}$, the ECM emission power becomes almost independent of the plasma density at the source of the Alfv\'en waves according to Eq.~\eqref{secm}. {However, such a regime corresponds to a super-Alfv\'enic interaction where a magnetohydrodynamic shock develops close to the planet and the Alfv\'en waves cannot propagate to the star making the excitation of auroral ECM emission at the star unfeasible.} Finally, we notice that the expression for the Alfv\'en wing power adopted by \citet{Noyolaetal14} and \citet{Narangetal23} is smaller by a factor of 2 with respect to that proposed by \citet{Sauretal13} (cf. their Eq. 55) in the limit $V_{0}/V_{\rm A} \rightarrow 0$, where $V_{\rm A} \equiv B_{\rm m}/\sqrt{\mu \rho_{\rm m}}$ is the Alfv\'en velocity. This was due to the consideration by \citet{Sauretal13} of the energy flux both inside and outside the flux tube of cross section $\pi R_{\rm m}^{2}$ that was not accounted for in the original work by \citet{Neubauer80}, according to \citet{Callinghmetal24} -- see their Eq. (3) and the related discussion.

A relevant point is that Eq.~\eqref{secm} provides a density estimate  at the source of the Alfv\'en waves, that is, close to the planet. In other words, it provides a snapshot that depends on the position of the planet along its orbit and can vary in time. 
Moreover, if the magnetic axis of the stellar dipole is inclined to the orbital angular momentum of the planet, $B_{\rm m}$ varies up to a factor of 2 at a given orbital radius, producing a variation in the estimated density up to a factor of 4 when $\rho \ll \rho_{\rm crit}$, that is, in the regime where the application of Eq.~\eqref{secm} is usually made. 

The density close to the planet is the highest in the magnetosphere because the planet is  the source of plasma (cf. Sect.~\ref{planet_mass_loss_mechanisms}). 
Therefore, the density derived from Eq.~\eqref{secm} can be used to estimate an upper limit to the plasma mass flow  
 due to magnetospheric corotating convection. 
 The maximum radial velocity of the magnetospheric convection at a distance $r$ from the center of the star is provided by \citep[][cf. their Eq.~26]{Hilletal81}
\begin{equation}
    \varv_{\rm mc} (r) \la \frac{1}{\sqrt{6}} \Omega_{\rm s} r \left( \frac{r}{R_{\rm H}}\right)^{2},
    \label{v_mconv}
\end{equation}
where $R_{\rm H}$ is the Hill radius (see Eq.~\ref{r_hill}).
Therefore,  in stationary conditions, an upper limit to the mass flow rate  can be estimated from the equation of mass continuity by integrating our upper limit to the mass flux $\rho_{\rm m}(a)\, \varv_{\rm mc}(a)$ over the lateral  surface of the cylinder of radius $a$ across which the plasma flows as a consequence of magnetospheric convection. In this way, our estimate of the upper limit to the mass flow, based on the density derived from   our star--planet interaction model, $\dot{M}_{\rm SPI}$, is 
\begin{equation}
    \dot{M}_{\rm SPI} \la  4\pi a  H_{\rm c} \, \rho_{\rm m}(a) \, \varv_{\rm mc}(a),
    \label{mdot_spi}
\end{equation}
where $a$ is the orbital distance of the planet, whose orbit is assumed circular,  $4\pi a H_{\rm c}$ is the lateral surface of the cylinder across which the mass flux  occurs, the density $\rho_{\rm m}(a)$ at distance $a$ is obtained from Eq.~\eqref{secm}, and the magnetospheric convective velocity $\varv_{\rm mc}(a)$ by Eq.~\eqref{v_mconv} with $r=a$. The semi-height of the cylinder, $H_{\rm c}$, is the so-called  centrifugal scale height, that is, the distance from the equator within which the outward flow, transported by magnetospheric convection, is largely confined. It is given by
\begin{equation}
H_{\rm c} = \left( \frac{2 k_{\rm b} T}{3 m\Omega_{\rm s}^{2}}\right)^{1/2},
\label{cen_hp}
\end{equation}
where $k_{\rm b}$ is the Boltzmann constant, $T$ the temperature of the magnetospheric plasma,  and $m$ the mass of the ions that comprise most of the mass of the magnetospheric plasma  \citep{Hilletal81}.  The plasma is assumed to corotate with the star because the Hill radius is on the order of $\sim 100$ stellar radii or larger in our case (see Sect.~\ref{res_wind_flux}). If that is not the case, the $\Omega_{\rm s}$ in Eq.~\eqref{cen_hp} should be replaced by the angular velocity of the plasma. 

The ion mass is computed considering a plasma composed of SO$_{2}$ ions, like in the case of  the Jupiter moon Io (cf.  Sect.~\ref{tidal_heating_mass_loss}), while the plasma has an adopted temperature that is given by the equilibrium temperature of a black body heated by the stellar radiation at distance $a$, that is, $T \la  T_{\rm eq} \equiv T_{\rm eff}\sqrt{R_{\rm s}/a}$, where $R_{\rm s}$ is the radius and $T_{\rm eff}$ the  effective temperature of the star (pickup ions can be remarkably hotter in some systems making $H_{\rm c} \propto T^{1/2}$ correspondingly greater). Equation~\eqref{mdot_spi} does not take into account the return flow toward the star that develops in the magnetospheric corotating convection; therefore, it provides us with an upper limit to the mass-loss rate $\dot{M}_{\rm SPI}$ moving outward across the magnetosphere.

Our simplified analysis neglects local plasma sources and sinks and assumes only an axisymmetric source in the inner region of the magnetosphere, so that the mass flow rate $\dot{M}$ is a constant and is independent of the radial coordinate. In the Hill model, the mass flow ultimately depends on the radial gradient in the plasma angular momentum per unit mass \citep[cf. Eq. 16 in][]{Turnpenneyetal17}. In a stationary regime, that gradient adjusts itself in order to transport the entire mass flow produced by the source of plasma in the inner magnetosphere. 
In our formulation, we aim at estimating an upper limit to such a mass flow. As displayed on  the r.h.s. of Eq.~\eqref{mdot_spi}, such an upper limit is proportional to $\varv_{\rm mc}(a)$. In turn, $\varv_{\rm mc} \propto R_{\rm H}^{-2} \propto \dot{M}^{1/2}$ from Eqs.~\eqref{v_mconv} and~\eqref{r_hill}. Therefore, in our approach, it is the actual value of the average density $\rho_{\rm m}(a)$ that is predicted to adjust itself in order to transport the required mass-loss rate across the cylindrical sheet of radius $a$. Such an adjustment is possible through a constant multiplicative factor because $\dot{M}$ is a constant independent of the radial distance in the Hill model.

\subsection{Magnetospheric mass flux due to an  ionospheric outflow}
\label{stellar_wind_main}

Ultra-cool and brown dwarf stars show X-ray emission, indicating the presence of a plasma with temperatures on the order of million kelvin in their outer atmospheres. Given the measured emission measure values and assuming that the hot plasma is confined by the magnetic field, it is possible to estimate its spatial extent finding that it is confined to $\la 1-10$\% of the stellar radius \citep[see Sect.~3.1.3 and the final paragraphs of Sect.~4.2 of][for details]{Magauddaetal24}. In other words, the coronae of UCD stars are geometrically thin. If a fraction of the coronal  magnetic field lines are open, for example as a result of magnetic reconnection processes, some part of such a hot plasma can flow into the surrounding ionosphere of the star being accelerated by its own thermal pressure and the centrifugal force produced by the rapid rotation of the star. The induced  mass flux across the ionosphere and  the magnetosphere is expected to be low because the plasma density decreases rapidly away from the surface of the star given its observed small emission measure. 

We introduce a model of such an ionospheric plasma outflow in Appendix~\ref{stellar_wind}, showing that its density decreases away from the star up to the corotation radius. The outflow  is approximately similar to that coming from Io in Jupiter's magnetosphere, that is, originating not far from the corotation radius in the inner magnetosphere of the planet. For the sake of simplicity, we treated the corresponding mass and angular momentum transport inside the magnetosphere of our star by means of the Hill model. This model  assumes that the plasma source is located at the base of the magnetosphere below or at the corotation radius and that mass is transported across the magnetosphere at a constant rate, independent of the radial distance (see Appendix~\ref{stellar_wind} for a discussion of such an assumption). 

In Appendix~\ref{stellar_wind}, we show that an upper limit to the mass flux across the magnetosphere of a UCD star is provided by Eq.~\eqref{wind_mass_loss_fin}. That upper limit is useful for our comparison with the mass flux as derived from the radio observations (cf. Sect.~\ref{res_wind_flux}). The main uncertainty in the  application  to LSR~J1835+3259 comes from the effective Pedersen conductance. Therefore, following \citet{Turnpenneyetal17}, we varied such a parameter between $0.01$ and 10~mho.

\subsection{Mechanisms producing mass loss from a putative planet}
\label{planet_mass_loss_mechanisms}

In addition to an ionospheric outflow, we considered the possibility of a putative planet orbiting our UCD star as a source of plasma to its magnetosphere. Its evaporation can be powered by tidal heating (Sects.~\ref{tidal_heating_mass_loss} and~\ref{ecc_maintaining}) and/or electric currents induced by its orbital motion inside the stellar magnetosphere (Sect.~\ref{induction_model_0} and Appendixes~\ref{induction_model} and~\ref{electric_conduc}).

\subsubsection{Tidal heating and the mass-loss rate}
\label{tidal_heating_mass_loss}
The source of ions and most of the electrons permeating Jupiter's magnetosphere is Io's volcanic activity, the energy of which is provided by tidal heating. With an observed Io surface mean heat flux of $\sim 2$~W~m$^{-2}$ \citep[e.g.,][]{Veederetal94,Tyleretal15}, its mean internal heating power is  $\sim 8.3 \times 10^{13}$~W. Such a  heating is produced by the tides induced by Jupiter on Io  whose orbit is slightly eccentric ($e \sim 0.0041$) owing to the perturbations by Europa and Ganymede. 

The power  dissipated by tides, $P_{\rm tide}$, for small eccentricity ($e \la 0.2$) is given by \citep[e.g.,][]{Jacksonetal08a,Milleretal09}
\begin{equation}
P_{\rm tide} = \frac{63}{4} \left[ \left( GM_{\rm s} \right)^{3/2} \left( \frac{M_{\rm s} R_{\rm m}^{5} e^{2}}{Q_{\rm m}^{\prime}}\right)\right] a^{-15/2}, 
\label{tida_diss}
\end{equation}
where $G$ is the gravitational constant, $M_{\rm s}$ the mass of the star, $R_{\rm m}$ the radius of the planet, $Q_{\rm m}^{\prime}$ the modified tidal quality factor of the planet  that quantifies the efficiency of tidal dissipation in its interior, and $a$ the orbit semimajor axis. The lower the modified quality factor $Q^{\prime}_{\rm m}$, the stronger the tidal dissipation. It is defined as $Q^{\prime}_{\rm m}= (3/2) (Q_{\rm m}/k_{2})$, where $Q_{\rm m}$ is the tidal quality factor and $k_{2}$ the tidal Love number of the planet. For a rocky body such as Io, $10^{2} \la Q^{\prime}_{\rm m} \la 10^{3}$ depending on its rheology and the tidal frequency  \citep[see the discussion in, e.g., Sect.~2.5 of][]{Lanza21}. Specifically in the case of Io, $Q^{\prime}_{\rm m}$ is about 100 according to \citet{Laineyetal09} that was confirmed by a recent determination based on  Juno Doppler measurements \citep{Parketal24}. 

The mass loss of our putative planet  occurs from its atmosphere. In the case of Io, its replenishing of  Jupiter magnetosphere is a complex process that includes  interactions between the upper atmosphere and the exosphere of the moon, the neutral clouds, the plasma torus,  the plasma sheet, and the magnetosphere of Jupiter with several key aspects not understood yet  \citep[see][]{Rothetal24}. However, the leading explanation of its mass loss involves atmospheric sputtering, i.e., momentum transfer between incoming ions and the upper volcanic atmosphere of Io \citep{Haffetal81,Johnson04}. The mass loss can be approximated as a function of plasma pressure by scaling to Io, as described in \citet{GebekOza20}, Eq.~(14):
\begin{equation}
    \dot{M}_{\rm sputtering} \sim  \frac{PV}{U} R_{\rm x}^{2} \dot{M}_{\rm Io}, 
\end{equation}
where $PV$ is the product of the plasma pressure and ion velocity, $U$ the binding energy, $R_{\rm x}$ the exobase radius, all of the above quantities scaled to Io's whose total mass-loss rate is $\dot{M}_{\rm Io}$. To estimate the sputtering of a trace volatile $i$, in the bulk atmosphere, one can add a prefactor of $x_{\rm i}$,  the mass mixing ratio of the volatile with respect to a nominal SO$_2$ background. At Io, described in \citet{Ozaetal19} Eq.~(7), this can be as large as $x_{\rm Na} \approx 0.1$ for sodium, at the exobase. 

We focused on the process that extracts neutral SO$_{2}$ from the exosphere of Io to ultimately fill in the plasma torus and assume that the molecules are ionized immediately after their extraction and diffuse through Jupiter's  magnetosphere. The dominant  process that extracts neutrals from Io is represented by the elastic collisions between the ions of the plasma torus, which corotates with Jupiter's magnetic field and overtakes Io with a velocity of 57 km~s$^{-1}$, and the SO$_{2}$ molecules in the exosphere of the moon. Such a sputtering process imparts velocities in excess of the escape velocity from Io of 2.56 km~s$^{-1}$ while dissociating and ionizing molecules, so that they leave the moon's atmosphere as ions. The mass-loss rate varies linearly with the atmospheric density at the surface of Io according to a  model by \citet{Saur03}. On the other hand, the surface atmospheric density depends on the gas outflows produced by active volcanos as well as on the sublimation of frozen SO$_{2}$ deposited on the surface by previous volcanic activity. Both processes lead to an atmospheric density that can be assumed proportional to the mass rate erupted by Io's volcanos averaged over a sufficiently long period of time, a quantity that can be assumed to be proportional to $P_{\rm tide}$. 

We assumed that similar processes occur in the case of a small rocky planet orbiting a UCD star and that the density of the plasma close to the planet surface is proportional to the mass-loss rate from its exosphere. Sulfur dioxide has now been identified by JWST/MIRI and may be an infrared satellite signature \citep{Oza2026} of the toroidal plasma processes we describe here. In this way, adopting the \citet{Saur03} model,  the density of the magnetospheric plasma close to the planet  can be assumed to be proportional to its average volcanically erupted mass rate and, therefore, to $P_{\rm tide}$, under stationary conditions. 
Since the mass flow rate transported by magnetospheric convection is proportional to the magnetospheric density $\rho(a)$ at the orbital radius $a$ of the planet (see, e.g., Eq.~\ref{mdot_spi}), the mass flow rate will be proportional to the tidal heating power, $P_{\rm tide}$, inside the volcanically active planet.

\subsubsection{Maintaining the eccentricity of the planetary orbit}
\label{ecc_maintaining}
{In the absence of further bodies in the system that excite it, the eccentricity of the orbit of the putative planet decays on an e-folding timescale given by
 Eq.~(4) by \citet{Jacksonetal08}, which we recast as
\begin{equation}
\tau_{\rm e} \equiv - \left( \frac{1}{e} \frac{de}{dt} \right)^{-1} =   \frac{2}{63\pi} P_{\rm orb} \, Q^{\prime}_{\rm m} \left( \frac{M_{\rm m}}{M_{\rm s}}\right) \left(\frac{a}{R_{\rm m}} \right)^{5}.  
\end{equation}
Here $P_{\rm orb}$ is the orbital period of the planet; the other symbols have already been introduced. For example, adopting $Q^{\prime}_{\rm m} \sim 100$, a semimajor axis $a=9.5$~$R_{\rm s}$, a planet  with the radius and mass of Io, and a star with the mass of LSR~J1835+3259,  we obtain a timescale for the decay of the eccentricity of only $\sim 7800$ years, that is, $\sim 20$ times shorter than the Io eccentricity damping timescale for the same $Q^{\prime}_{\rm m}$. 

One of the simplest scenarios for keeping the orbit of the planet eccentric is the presence of a second planet on a coplanar, nonresonant, external orbit as in the simple model by \citet{Mardling07}. In such a scenario, the orbit of the inner planet can be maintained at a small almost constant eccentricity value ($e \approx 10^{-3}$) for a timescale of several gigayears. Such an "equilibrium" eccentricity is given by Eq. (36) of \citet{Mardling07}, while the timescale of its very slow decay can be computed with the method presented in Sect.~3.3.3 of the same paper. {In our implementation of Eq. (36) of \citet{Mardling07}, in addition to the general relativity precession, we considered in their Eq.~(34) the contribution of the rotational deformation of the star to the apsidal precession to be \citep[see Eqs.~31 and~49 of][]{MardlingLin02}
\begin{equation}
    \dot{\eta}_{\rm rot} = \frac{k_{\rm s}}{n_{\rm p}} \frac{(1+M_{\rm m}/M_{\rm s})}{(1-e^{2})^{2}} \left(\frac{R_{\rm s}}{a} \right)^{5}\Omega_{\rm s}^{2},
\end{equation}
where $k_{\rm s}=0.14$ is the apsidal motion constant for a fully convective star, $n_{\rm p}= 2\pi/P_{\rm orb}$ is the mean orbital motion of the planet, and the stellar spin is aligned with the orbital angular momentum. 

We adopted the \citet{Mardling07} model to justify maintaining of the eccentricity of the inner putative planet, the volcanic activity of which provides plasma to the magnetosphere of our UCD star. We apply the model in Sect.~\ref{res_spi_flux} to estimate the equilibrium eccentricity, the tidal power heating rate, and the long timescale of its decay. 
}}

\subsubsection{Induction heating}
\label{induction_model_0}
The strong magnetic fields of UCDs and BDs make  electromagnetic induction heating possible in the interiors of their close-by planets. As we shall see in Sect.~\ref{induction_heat_sect}, such a mechanism generally does not provide enough power to justify the required evaporation rates from those planets according to the model presented in Appendixes~\ref{induction_model} and~\ref{electric_conduc}.

\section{Application to LSR J1835+3259}
\label{application}

The main parameters of LSR~J1835+3259  are listed in Table~\ref{table1} as reported by \citet{Kaoetal23}. The magnetic field intensity at the pole $B_{\rm p}$ is estimated from the ECM emission frequency as discussed in Sect.~\ref{intro}. 
In this section, first we apply the models presented in Sects.~\ref{ionospheric-magnetospheric} (magnetospheric--ionospheric coupling or M-I model) and~\ref{density_from_ECM} (Alfv\'en wing model) to estimate the mass flow across the stellar magnetosphere from its observed ECM radio emission (see Sects.~\ref{mdot_ionospheric-magnetospheric} and~\ref{res_spi_flux}). In Table~\ref{table_comp}, we report  the estimates of the upper limits of the mass flow rate across the magnetosphere, $\dot{M}$, according to the M-I and Alfv\'en wing models, the latter for four values of the effective Pedersen conductance, $\Sigma_{\rm p}^{*}$. We also list the results of a comparison of such values with the theoretical mass flow rates from a stellar ionospheric outflow and from a putative planet orbiting around our UCD star with internal tidal heating and electric induction heating  as predicted by the models in Appendix~\ref{stellar_wind} and Sect.~\ref{planet_mass_loss_mechanisms} (see Sects.~\ref{res_wind_flux}--\ref{induction_heat_sect}). Details on the determination of the mass-loss rates from the ECM flux or on its theoretical estimates are provided below in the respective subsections.

\begin{table}
\begin{center}
\caption{Parameters of the UCD star LSR~J1835+3259 }
\begin{tabular}{lr}
\hline
& \\
Parameter & Value\\
 & \\
 \hline
  & \\
Mass (M$_{\rm J})$ & $77.28 \pm 10.34$  \\
Radius (R$_{\rm J}$) & $1.07 \pm 0.05$ \\
$T_{\rm eff}$ (K) & $2800 \pm 30$ \\
Distance (pc) & $5.6875 \pm 0.00292$ \\
Spectral type & M8.5V \\
Rotation Period (hr) & $2.84 \pm 0.01$ \\
$B_{\rm p}$ (T) & $\sim 0.3$ \\
 & \\
\hline
\label{table1}
\end{tabular}
\end{center}
\end{table}

\begin{table*}
\caption{Summary of the comparison of the upper limits to the mass flow rates across the stellar magnetosphere as obtained from the ECM emission with the rates allowed by the different theoretical models. }
\begin{tabular}{|c||c|c|c|}
\hline
$\dot{M}$ from observed ECM emission & & Theoretical $\dot{M}$ from models & \\
\hline
& Ionospheric outflow & Planet tidal heating & Planet induction heating \\ 
\hline
& & $6 \leq a/R_{\rm s} \leq 10$, Io-like planet & \makecell{$ 6 \leq a/R_{\rm s} \leq 10$, Io-like planet, \\ 
$\langle \sigma \rangle = 0.1$~mho~m$^{-1}$}\\ 
\hline 
\makecell{M-I model, closed magnetosphere, \\ $\dot{M} \sim 10^{5}$~kg~s$^{-1}$} & \makecell{YES, \\  if $\Sigma^{*}_{\rm p} \la {0.003}$~mho} & YES & NO \\ 
\hline
\makecell{M-I model, open magnetosphere, \\ $\dot{M} \sim 10^{3}$~kg~s$^{-1}$} & \makecell{YES, \\ if $\Sigma^{*}_{\rm p} \la {0.3}$~mho} & YES & YES \\ 
\hline 
Alfv\'en Wing model, $ 6 \leq a/R_{\rm s} \leq 10$ &  & & \\
\hline
$\Sigma_{\rm p}^{*} = 10.0$~mho: $\dot{M} \sim (0.05 - 1.8) \times 10^{4}$~kg~s$^{-1}$ & NO & YES & YES \\ 
\hline 
$\Sigma_{\rm p}^{*} = 1.0$~mho: $\dot{M} \sim (0.05 - 1.8) \times 10^{5}$~kg~s$^{-1}$ & NO & YES & NO \\ 
\hline
$\Sigma_{\rm p}^{*} = 0.1$~mho: $\dot{M} \sim (0.05 - 1.8) \times 10^{6}$~kg~s$^{-1}$ & NO & YES & NO \\ 
\hline
$\Sigma_{\rm p}^{*} = 0.01$~mho: $\dot{M} \sim (0.05 - 1.8) \times 10^{7}$~kg~s$^{-1}$ & NO & NO & NO \\ 
\hline
\end{tabular}
\tablefoot{We report the results of the comparison between the measured and the theoretical values to conclude whether a given model is suitable to account for the estimated mass flux (yes) or not (no).}
\label{table_comp}
\end{table*}

\subsection{Auroral ovals: Estimating mass flux from magnetospheric--ionospheric coupling}
\label{mdot_ionospheric-magnetospheric}

\citet{Turnpenneyetal17} provide estimates of the mass  flux across the magnetosphere of a UCD star required to account for the observed ECM emission fluxes for the range of electrodynamic parameters considered in Sect.~\ref{ionospheric-magnetospheric}, a rotation period of $\sim 2$ hours, and a surface magnetic dipole field of 0.15~T at the equator. Such rotation period and magnetic field apply well to the case of LSR~J1835+3259. According to \citet{Hallinanetal08} and \citet{Kaoetal23}, the bursty emission, attributed to ECM, has  an integrated flux $S$ in the 8.4~GHz band of $\sim 2500 $~$\mu$Jy.

We refer the reader to Sect.~3.2 of \citet{Turnpenneyetal17} for a detailed account. In Table~\ref{table_comp}, we report  only the maximum values of the mass flux that are on the order of $\dot{M}\sim 10^{5}$ or  $\sim 10^{3}$~kg~s$^{-1}$ for a closed or an open magnetosphere with the UCD star moving with a velocity of 50~km~s$^{-1}$ with respect to the interstellar medium, respectively. We recall that in the \citet{Turnpenneyetal17} modeling the effective Pedersen conductance is assumed to vary between $0.01$ and $10.0$~mho. 

\subsection{Planetary footprints: Estimating mass flux from the star--planet interaction model }
\label{res_spi_flux}

Next we assumed that the ECM emission of LSR~J1835+3259 is powered by a star--planet interaction mediated by Alfv\'en waves excited by the planet moving across the stellar magnetosphere.  
The bursty emission flux was assumed to be $S =2500$~$\mu$Jy (as in Sect.~\ref{mdot_ionospheric-magnetospheric}).

\begin{figure}
\centerline{
\includegraphics[width=8cm,height=11cm,angle=270]{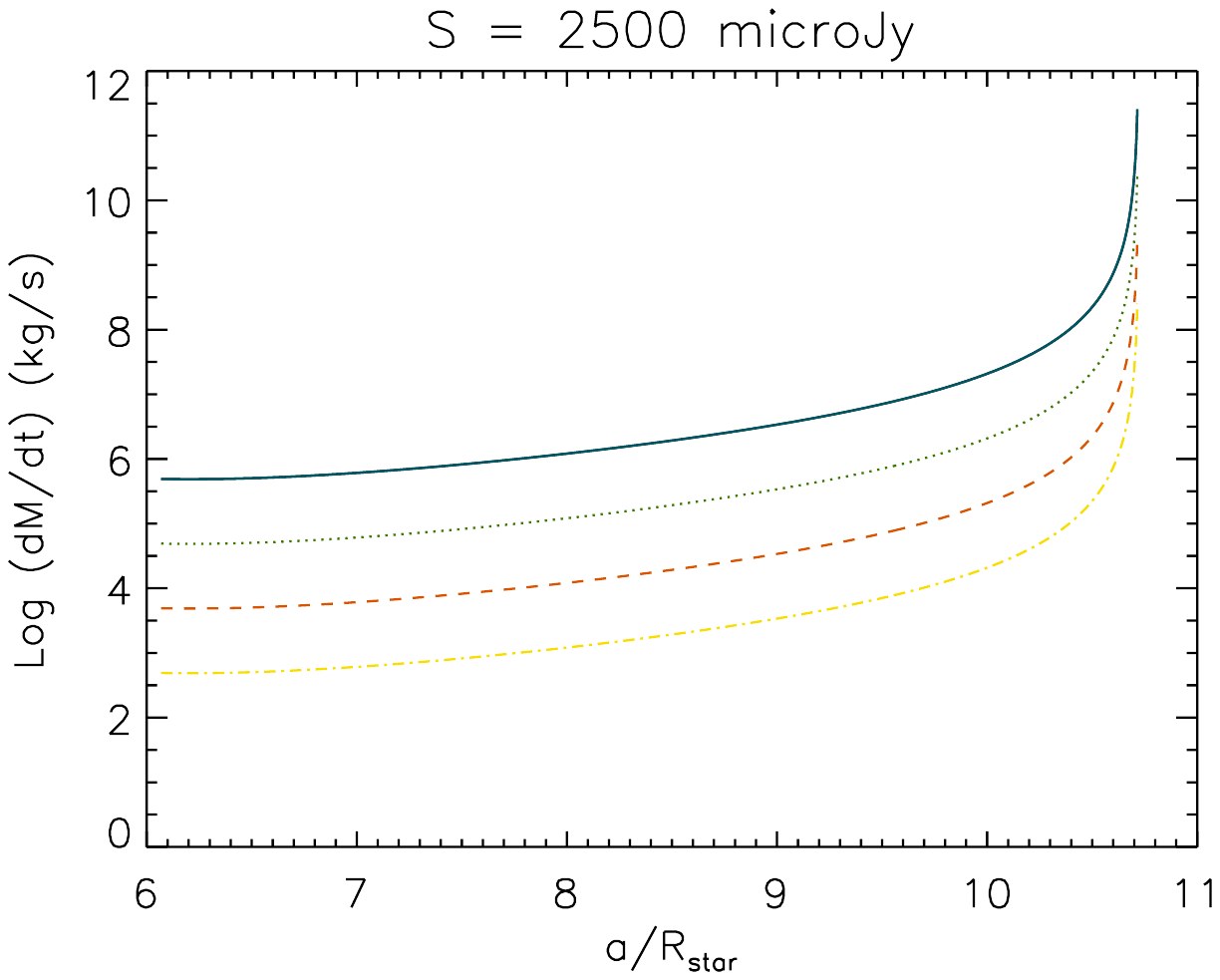}}
\caption{Upper limit to the mass flux $\dot{M}_{\rm SPI}$ across the magnetosphere of LSR~J1835+3259 vs. the orbital radius of a planet with the radius of Io as estimated by means of Eq.~\eqref{mdot_spi} and assuming the star--planet interaction and magnetospheric convection model  (see Sect.~\ref{density_from_ECM}) and an observed auroral radio flux $S= 2500$~$\mu$Jy. Different line colors and line styles refer to different values of the effective Pedersen conductance: $\Sigma_{\rm p}^{*} = 0.01$ mho (solid cyan); $\Sigma_{\rm p}^{*} = 0.1$ mho (dotted green); $\Sigma_{\rm p}^{*} = 1.0$ mho (dashed orange); and $\Sigma_{\rm p}^{*} = 10.0$ mho (dash-dotted yellow). }
\label{spi_density_fig1}
\end{figure}

The corresponding mass flow across the magnetosphere of the star, estimated with the model introduced in Sect.~\ref{density_from_ECM}, is shown in Fig.~\ref{spi_density_fig1}. We considered a planet with the same radius as Io orbiting outside the Roche limit -- within which a planet would be destroyed by the tidal force of the star ($\la 6$~$R_{\rm s}$) -- and plot the mass flow across the magnetosphere versus  the semimajor axis of the planetary orbit up to the distance where the magnetic field reaches the limit value in Eq.~\eqref{dist_div} and the density diverges ($a/R_{\rm s} \sim 10.7$ in the present case). We plot the mass flow for four different values of the effective Pedersen conductance $\Sigma_{\rm p}^{*}$ in the range adopted by \citet{Turnpenneyetal17}. The plasma is assumed to consist of ionized SO$_{2}$ molecules as in the case of Io, giving a mass of 64.07 atomic mass units in Eq.~\eqref{cen_hp}, while the plasma temperature $T$  is the black body equilibrium temperature at the distance of the planet, computed with  the stellar effective temperature as listed in Table~\ref{table1} \citep[][]{Boehmetal25}. 

The mass flux in Fig.~\ref{spi_density_fig1}  increases  with the semimajor axis of the planet, $a$, for a fixed  Pedersen conductance, $\Sigma_{\rm p}^{*}$, because the density as given by Eq.~\eqref{secm} and the flow velocity  due to corotating magnetospheric convection, $\varv_{\rm mc}$, both increase  with $a$,  while the centrifugal scale height decreases slowly as $H_{\rm c} \propto a^{-1/4}$. 
To generalize our upper limit to the mass flux given by Eq.~\eqref{mdot_spi} to other values of the Pedersen conductance, we considered the dependence  of the different terms appearing in that equation on $\Sigma_{\rm p}^{*}$.
For a given density at the distance of the planet $\rho_{\rm m}(a)$ as given by Eq.~\eqref{secm} that does not depend on $\Sigma_{\rm p}^{*}$, equation~\eqref{mdot_spi} gives an upper limit to the mass flow $\dot{M}_{\rm SPI} \propto \varv_{\rm mc}$ because $H_{\rm c}$ does not depend on $\Sigma_{\rm p}^{*}$. In turn, $\varv_{\rm mc} \propto R_{\rm H}^{-2}$ from Eq.~\eqref{v_mconv} and $R_{\rm H} \propto (\Sigma^{*}_{\rm p}/\dot{M})^{1/4}$ from Eq.~\eqref{r_hill}, yielding the dependence of the mass flow rate on Pedersen conductance as ${\dot{M}_{\rm SPI} \propto  (\Sigma_{\rm p}^{*})^{-1}}$.   
The dependence of the mass flux on the radius of the planet is remarkably stronger with { ${\dot{M}_{\rm SPI} \propto [\rho_{\rm m}(a)]^{2} \propto R_{\rm m}^{-8}}$} because a larger cross section of the planet makes the excitation of the Alfv\'en waves that power the ECM emission more efficient (cf. Eq.~\ref{secm}).

In Table~\ref{table_comp}, we list the ranges of the upper limits of the mass flow rates as obtained for an Io-like planet orbiting between $6\, R_{\rm s}$ (the Roche limit) and $10\, R_{\rm s}$ stellar radii for four different values of the effective Pedersen conductance, $\Sigma_{\rm p}^{*}$. The lower values of the reported ranges refer to the lower limit of the orbit semimajor axis, $a = 6\, R_{\rm s}$, while the upper values refer to $a= 10\, R_{\rm s}$. For $a> 10\, R_{\rm s}$, the mass flow rate diverges rapidly reaching unrealistically large values and the model cannot be applied for $a > 10.7\, R_{\rm s}$ because the star--planet interaction becomes super-Alfv\'enic (cf. Sect.~\ref{density_from_ECM}). 
  
\subsection{Coronal escape: Theoretical mass flux from an ionospheric outflow}
\label{res_wind_flux}
LSR~J1835+3259 has an upper limit to its X-ray luminosity in the $0.1-2$~keV band of $3.3 \times 10^{17}$ W \citep{Bergeretal08}. Adopting a typical coronal temperature $T_{\rm cor}=1.5$~MK as in \citet{Schrijver09}, we obtain an upper limit for its coronal base density $n_{e0} = 1.51 \times 10^{14}$ m$^{-3}$ from Eq.~\eqref{corona_lum} that for a hydrogen plasma means $\rho_{0} = 2.52 \times 10^{-13}$~kg~m$^{-3}$. In the case of our star, the coronal pressure scale height is $H_{\rm p}=0.1624$~$R_{\rm s}$ for the above coronal temperature, while the corotation radius is $R_{\rm cor} = 3.87$~$R_{\rm s}$ from Eq.~\eqref{rcorot}. The centrifugal scale height is $H_{\rm c}=1.93$~$R_{\rm s}$ as given by Eq.~\eqref{cen_hp} for a  hydrogen plasma with a temperature of 1.5~MK. The fraction of the surface of the star covered by the magnetic field lines along which there is a  ionospheric outflow, $f_{\rm w}$, is taken as the mean value of the fractional area of the solar coronal holes as estimated by \citet{Soonetal00}, that is, $f_{\rm w}= 0.15$ in Eq.~\eqref{wind_mass_loss_fin}. 

The upper limit for the mass flow across the magnetosphere of our star, $\dot{M}_{\rm ion\, flow}$, which is supported by a ionospheric outflow  originating from the base of the corona according to the model presented in Appendix~\ref{stellar_wind}, is given in Table~\ref{table2} together with the upper limit for the magnetospheric convection velocity, $\varv_{\rm mc}$, and the Hill radius,  $R_{\rm H}$, for four values of the effective Pedersen conductance, $\Sigma_{\rm p}^{*}$, that span the range  adopted by \citet{Turnpenneyetal17}. 
In our model, the density at the level of the magnetosphere where corotating convection begins (that is, at $r=R_{\rm cor}$) is independent of the magnetospheric conductance,  $\Sigma_{\rm p}^{*}$, and depends only on the  ionospheric outflow model, that is, on the adopted  ionospheric outflow  temperature, $T_{\rm w}$, and base density, $\rho_{0}$, as well as the stellar mass, radius, and angular velocity (cf. Eq.~\ref{density_eq}). Therefore, the differences in the values of the mass flux listed in Table~\ref{table2}  depend on the differences in the velocity of the corotating magnetospheric  convection, $\varv_{\rm mc}$, which in turn depends on the differences in the Hill radius, $R_{\rm H}$,  through the differences in the adopted Pedersen conductance (cf. Eqs.~\ref{v_mconv} and~\ref{r_hill}). Specifically, $\varv_{\rm mc}$ decreases as $R_{\rm H}^{-2}$ with increasing Hill radius when the conductance increases. For that reason, the mass flux decreases with increasing Pedersen conductance. 
For the values of $\Sigma_{\rm p}^{*}$ not listed in Table~\ref{table2}, the mass flow can be  evaluated by means of  Eq.~\eqref{wind_mass_loss_fin}  that shows that  $\dot{M}_{\rm ion~flow} \propto (\Sigma_{\rm p}^{*})^{-1}$. 

Our upper limits for the mass flux supplied by an ionospheric outflow are sufficient to account for the ECM emission of LSR~J1835+3259 in the context of the ionospheric--magnetospheric coupling model and in the case of a closed magnetosphere, only if its ionospheric Pedersen effective conductance is sufficiently small, that is, $\Sigma_{\rm p}^{*} \la 0.003$~mho, because in that case the model by \citet{Turnpenneyetal17} requires $\dot{M}$ to be $ \sim 10^{5}$~kg~s$^{-1}$.  In their Fig.~10 (lower middle column), \citet{Turnpenneyetal17} demonstrate that a Pedersen conductance of 0.7~mho is capable of reproducing the observed radio flux with a mass flow rate of $\dot{M} \sim 10^{5}$~kg~s$^{-1}$ and the magnetic field of the star inferred from its radio emission. Our model indicates that, adopting such a conductance and the observed upper limit on the X-ray stellar luminosity, the physical process originating the mass flow across the magnetosphere cannot be an ionospheric outflow.

In the case of an open magnetosphere interacting with the interstellar medium, a larger conductance is allowed ($\Sigma_{\rm p}^{*} \la { 0.3}$ mho)  because a mass flow of only $10^{3}$~kg~s$^{-1}$ is enough to account for the observed ECM emission according to \citet{Turnpenneyetal17} (cf. Sect.~\ref{mdot_ionospheric-magnetospheric}).   {This value of $\dot{M}$ is adopted in Fig.~10 of \citet{Turnpenneyetal17}, who assume a Pedersen conductance of $1.2$~mho  (cf. their Fig.~10, middle upper panel).  Therefore, also in this case, an ionospheric outflow does not appear to be the process responsible for the mass flow across the stellar magnetosphere. }

If the ECM emission is powered by the Alfv\'en waves excited by a planet orbiting within $6-10$~$R_{\rm s}$, our  ionospheric outflow model is not capable of supplying a sufficient mass flux to account for the observed ECM flux  (cf. Table~\ref{table2} with Fig.~\ref{spi_density_fig1}). These results are reported in the second column from the left of Table~\ref{table_comp}.

\begin{table}
\begin{center}
\caption{Model parameters and upper limits to the magnetospheric mass flux for the stellar  ionospheric outflow model of the UCD star LSR~J1835+3259. }
\begin{tabular}{rrrr}
\hline
& & & \\
$\Sigma_{\rm p}^{*}$ & $\varv_{\rm mc}$ & $R_{\rm H}$ & $\dot{M}_{\rm ion\, flow}$ \\
(mho) & $\; ($m~s$^{-1}$) & ($R_{\rm s}$) & (kg~s$^{-1}$) \\
                                         \hline
               & & & \\ 
    0.01 &   185.04 &    77.50 & 3.43e+04 \\
    0.10 &    18.50 &   245.09 & 3.43e+03 \\
    1.00 &     1.85 &   775.03 & 3.43e+02 \\
   10.00 &     0.19 &  2450.86 & 3.43d+01 \\
 & & & \\
\hline
\label{table2}
\end{tabular}
\tablefoot{$\Sigma_{\rm p}^{*}$ is the magnetospheric conductance, $\varv_{\rm mc}$ the radial velocity of the magnetospheric convection (Eq.~\ref{v_mconv}), $R_{\rm H}$ the Hill radius (Eq.~\ref{r_hill}), and  $\dot{M}_{\rm ion\, flow}$ the stellar ionospheric outflow mass-loss rate (Eq.~\ref{wind_mass_loss_fin}) (cf. Sect.~\ref{stellar_wind_main} and Appendix~\ref{stellar_wind} for model details).  }
\end{center}
\end{table}

\subsection{Volcanic planet: Theoretical mass flux from tidal heating}
\label{res_tidal_heating}

\begin{figure}
\centerline{
\includegraphics[width=10cm,height=12cm,angle=0]{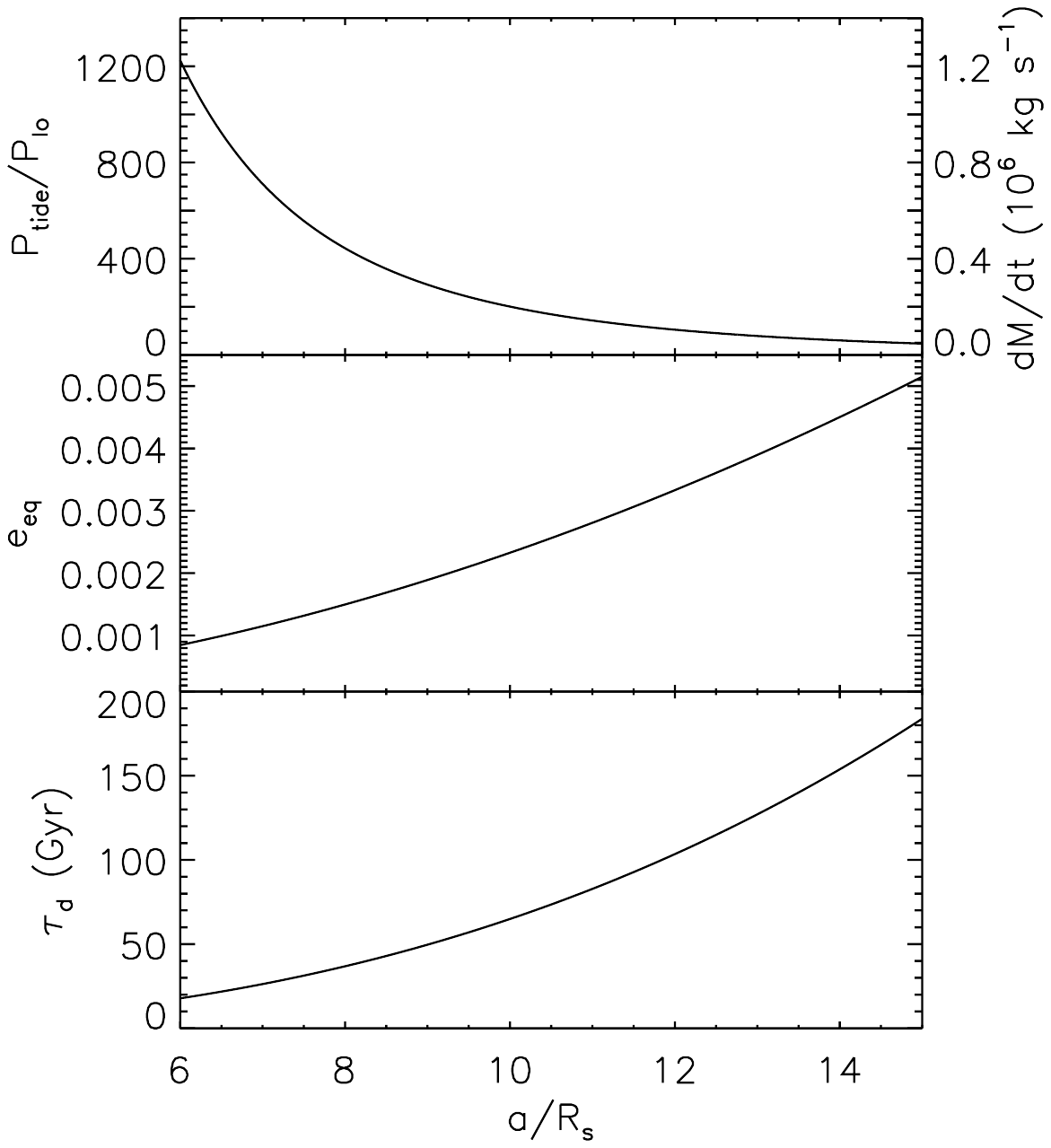}}
\vspace*{-1.0cm}
\caption{{\em Top}: Ratio of the tidal power dissipated in the putative inner planet to the tidal power dissipated inside Io (on the left axis) and mass-loss rate (on the right axis) vs. the orbit semimajor axis of the planet.  {\em Middle}: Equilibrium eccentricity of the orbit of the inner planet vs. the semimajor axis of its orbit. {\em Bottom}: Timescale for the decay of the equilibrium eccentricity of the inner planet and of the eccentricity of the outer planet.}
\label{tidal_quant}
\end{figure}

Next we considered the mass flux across the stellar magnetosphere that can be supplied by the volcanic activity of a small planet with the radius and mass of Io orbiting at different distances from LSR~J1835+3259. The mass flux can be assumed to be proportional to the tidal power dissipated inside the planet following  the discussion in Sect.~\ref{tidal_heating_mass_loss}. Therefore, in the top panel of Fig.~\ref{tidal_quant}, we plot on the left axis the ratio of the tidal power dissipated inside the  planet, $P_{\rm tide}$, to that dissipated inside Io, $P_{\rm Io}$, versus the semimajor axis, $a$, of the planetary orbit. The averaged mass-loss rate of Io is $\sim 10^{3}$~kg~s$^{-1}$; hence, the corresponding mass flux supplied by the planet activity is 
$\dot{M} \sim 10^{3} (P_{\rm tide}/P_{\rm Io})$~kg~s$^{-1}$ and is reported on the right axis of Fig.~\ref{tidal_quant}.
The tidal power, $P_{\rm tide}$, was computed by means of Eq.~\eqref{tida_diss} with the equilibrium eccentricity, $e^{\rm (eq)}_{\rm p}$, as given by Eq.~(36) of \citet{Mardling07}, assuming a second coplanar nonresonant planet in the system with an orbit semimajor axis $a_{\rm c}= 3.5\, a$ (where $a$ is the semimajor axis of the volcanically active inner planet), a mass $m_{\rm c}=0.4$~M$_{\oplus}$, and an orbit eccentricity $e_{\rm c}=0.5$. A perturbing planet on a resonant orbit as in the case of Io, appears to be a less likely scenario according to \citet{Cillibrasietal21}. The modified tidal quality factor $Q^{\prime}_{\rm m}=100$, which  is, equal to that measured in the case of Io, while  the mass of the star is  that of LSR~J1835+3259 in Table~\ref{table1}. 

The equilibrium eccentricity $e^{\rm (eq)}_{\rm p}$ is plotted in the middle panel of Fig.~\ref{tidal_quant} versus $a$. For comparison, the forced eccentricity of the orbit of Io is 0.0041, a value comparable to those obtained for the inner active planet in our putative system. In the bottom panel of Fig.~\ref{tidal_quant}, we plot the timescale for the decay of the equilibrium eccentricity, which gives the duration of the planet volcanic activity in the system. The decay timescale is longer than the Hubble time, indicating that the equilibrium eccentricity is maintained throughout the evolution of the system. 

A mass flux of $10^{5}$~kg~s$^{-1}$ can be supplied by an evaporating active planet with the above parameters up to a distance of $\sim 13$~$R_{\rm s}$, despite the rapid decrease in tidal dissipation with increasing semimajor axis. This is sufficient for the mass flux required by the ionospheric--magnetospheric coupling model for the ECM radio emission  by \citet{Turnpenneyetal17}. 

According to our model, an Io-like planet can reach a  mass flux rate up to $\dot{M} \sim 1.2 \times 10^{6}$~kg~s$^{-1}$ close to the Roche limit at $a \sim 6\, R_{\rm s}$ (cf. Fig.~\ref{tidal_quant}). If the ECM emission is powered by the Alfv\'en waves due to a star--planet interaction, such a maximum mass-loss rate is sufficient to account for the estimated mass flux across the magnetosphere, except for the lowest value of the effective Pedersen conductance ($\Sigma_{\rm p}^{*} \la 0.01$~mho) and $a \ga 9\, R_{\rm s}$ (cf. Fig.~\ref{spi_density_fig1}). These results are   summarized in Table~\ref{table_comp}. 

Finally, we note that the large mass flux of $10^{6}-10^{7}$~kg~s$^{-1}$ required in the case of a low effective Pedersen conductance  ($\Sigma_{\rm p}^{*} \la 0.01$~mho, cf. Fig.~\ref{spi_density_fig1}) could be supplied by a rocky planet with a radius larger by a factor of $1.6-2.5$ than the radius of Io and with the same value of $Q^{\prime}_{\rm m}$, given the strong dependence of $P_{\rm tide} \propto R_{\rm m}^{5}$ in Eq.~\eqref{tida_diss}. Planets with a radius comparable to that of the Earth have indeed been found around the UCD star TRAPPIST-1, but here we focus on an Io-like planet because the $Q^{\prime}_{\rm m}$ of Io has been estimated, while the rheological properties of those bodies are unknown. 

Having demonstrated that a specific two-planet system can provide a sufficient mass flux to the magnetosphere of LSR~J1835+3259, we do not explore two-planet systems with different  $a_{\rm c}/a, e_{\rm c}$, and $m_{\rm c}$ for the second outer planet. Such an exploration would be greatly time consuming and would not change our  conclusion.

\subsection{Induced electric currents: Theoretical mass flux from induction heating}
\label{induction_heat_sect}

An additional source of internal heating is provided by induced electric currents as discussed in Appendix~\ref{induction_model}. By assuming an orbital radius $a=9.2$~$R_{\rm s}$ and an average internal conductivity of $\langle \sigma \rangle \sim 0.01$~mho~m$^{-1}$ for a planet with the radius of Io whose orbit is inclined by $I=35^{\circ}$ to the magnetic equatorial plane of the UCD star, we find a skin depth $\delta \sim 0.32\, R_{\rm m}$ and a dissipated power of $\sim 7.7 \times 10^{13}$~W, which is comparable with the heating rate of Io. This is clearly insufficient to account for the mass flux required by the ionospheric--magnetospheric coupling model, except for an open magnetosphere (see Sect.~\ref{mdot_ionospheric-magnetospheric}),  or by the star--planet interaction model (see  Sect.~\ref{res_spi_flux}).
Even adopting a conductivity of $\langle \sigma \rangle \sim 0.1$~mho~m$^{-1}$, we obtain a skin depth, $\delta$, of $ \sim 0.1 \, R_{\rm m}$ and a power of $\sim 3.7 \times 10^{14}$~W, i.e., a factor of $\sim 4.7$ times that of Io, which is clearly not enough. Only if we assume that the outer 2\% of the radius of the planet consists of a global salty ocean with a mean conductivity $\langle \sigma \rangle$ of $  \sim 3.5$~mho~m$^{-1}$ (i.e., similar to that of the Earth's oceans) do we have a dissipated power of $\sim 2.6 \times 10^{15}$~W (which is $\sim 33$ times that of Io); in this case, the skin depth, $\delta$, is comparable with the depth of the ocean. However, the equilibrium temperature of a putative planet located at, say, $10$ Jupiter radii from our UCD star would be  $\sim 510$~K at the substellar point, even assuming a Bond albedo of 0.9. Such a temperature would rule out the presence of a liquid ocean on the hemisphere facing the star. On the opposite hemisphere, any water ocean would be always frozen owing to the tidal synchronization of planetary rotation. Given the much lower electric conductivity of the ice \citep{Okadaetal14}, the induction heating would be insufficient inside it.

We report our conclusions on the induction heating in Table~\ref{table_comp}, considering a planet with an orbit semimajor axis $6 \leq a \leq 10$~$R_{\rm s}$. We see that this process is not capable of providing a sufficient heating in most of the cases. In other words, tidal heating is expected to be dominant in the case of an Io-like planet in that range of separation from the star.

\begin{figure*}[ht!]
    \centering
    \includegraphics[width=\textwidth]{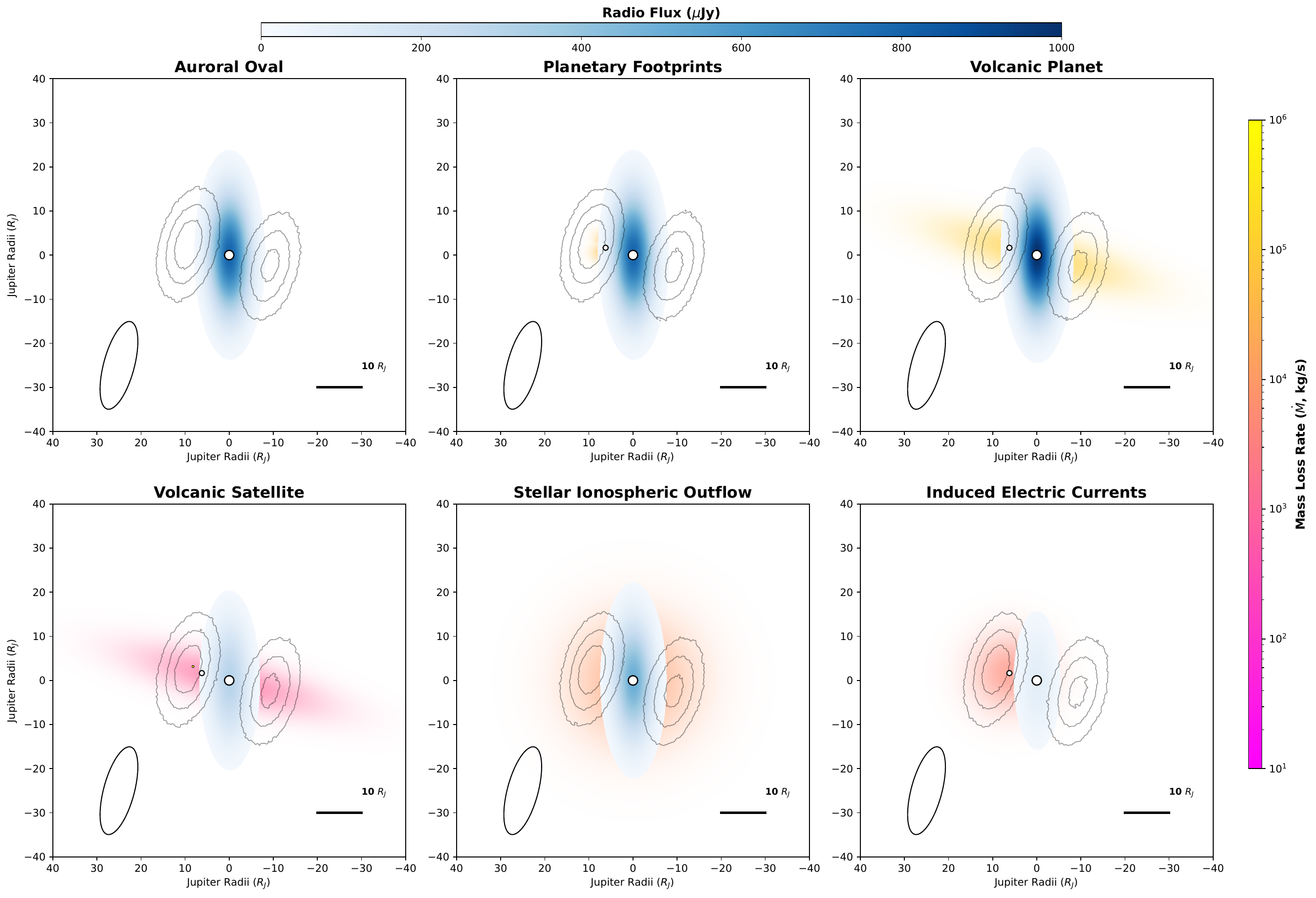}%{Fig3_lsr_j1835_80rj_beamfix_2026-09-09.png}
    \caption{Schematic maps of the microwave emission assuming different plasma sources for the magnetospheric environment of LSR J1835+3259 modeled over a $80$~$R_{\rm J}\times 80$~$R_{\rm J}$ field of view. 
    Six radio flux maps are simulated at 8.4~GHz; they follow the six plasma source processes presented as toy models in Section~\ref{model} and applied in Section~\ref{application}:  
    (1) auroral oval, (2) planetary footprints, (3) volcanic planet, (4) volcanic satellite, 
    (5) stellar ionospheric outflow, and (6) heating by induced electric currents. The synthesized radio observation beam is shown in the lower-left corner. 
    In all panels, the quiescent radiation emission (isocontours in black) has been added and  is depicted as symmetric lobes at approximately 9~$R_{\star}$ ($\sim 9.63$~R$_{\rm J}$). Mass-loss rates ($\dot{M}$) from Table 2 are shown for various dipolar and toroidal geometries. 
    The resultant auroral emission (in blue) peaks at an integrated flux of less than $\sim 1000~\mu\text{Jy}$. It is convolved with the Epoch~2 synthesized beam of \citet{Kaoetal23} ($1.71 \times 0.58$~mas) to approximate very long-baseline interferometry observations with a $16$~R$_{\rm J}$ auroral smudge (see Appendix~\ref{app:figure_construction} for details). 
    The central white circle indicates the UCD (radius $1.07$~R$_{\rm J}$) and an Io-size companion is placed at an orbit of $\sim 10$~R$_{\rm J}$  (smaller white circle) in the lower-left panel (volcanic satellite).}
    \label{fig:magnetosphere_sims}
\end{figure*}

\section{Discussion}
\label{discussion}

We estimated upper limits for the mass flux across the magnetosphere of the UCD star LSR~J1835+3259 considering its ECM emission.  We have compared those upper limits with the mass fluxes computed for different plasma sources and the radial transport by corotating magnetospheric convection due to the rapid rotation of this strongly magnetic star  (see Table~\ref{table_comp} for a summary of our results). 

Figure~\ref{fig:magnetosphere_sims} shows the simulated ECM radio emission for different  mass-loss rates, $\dot{M}$, of the planet (or satellite) according to the different proposed mechanisms in Sects.~\ref{model} and~\ref{application}, to provide some illustrative applications of our models -- see Appendix~\ref{app:figure_construction} for details on the method for simulating the emission flux and geometry of the very long-baseline interferometry observations, and the adopted mass-loss rates. The synchrotron quiescent radio emission as provided by the observations is added for completeness. We  cannot simulate it, given the uncertainties on the mechanisms that accelerate and transport relativistic electrons (see Sect.~\ref{intro}); therefore, we extracted information on the geometry and fluxes from the observations by \citet{Kaoetal23} (see Appendix~\ref{app:figure_construction} for details). In Fig.~\ref{fig:magnetosphere_sims}, only the volcanic planet scenario yields ECM fluxes on the order of $10^{3}$~$\mu$Jy that can account for the observed fluxes  (cf.~Sect.~\ref{intro}). A similar model was proposed by \citet{Boehmetal25} to account for the mass budget of circumstellar plasma tori produced by volcanically active close-by planets, while \citet{Oza2026}, by modeling neutral SO$_2$,  identified a candidate volcanic exomoon in the system of WASP-39~b by means of observations with JWST/MIRI. 

Our estimates of the stellar ionospheric outflow fluxes (see Table~\ref{table_comp}) are based on an analytical model rather than the empirical scaling relationships derived from a few indirect observations of mass-loss rates in the winds of stars of earlier spectral types \citep{Vidotto21}. We refrain from applying these empirical relationships because the coronae of UCD stars do not follow the empirical G\"udel-Benz law and are probably remarkably different from those of earlier-type stars \citep{Magauddaetal24}. 

In view of the limitation of the stellar ionospheric outflow model, we considered a volcanically active planet that can be internally heated by tides or by electric induction: we find the former to be the dominant process. 
Considering an orbit semimajor axis between 6 and 10 stellar radii, the planetary transit probability is between 10\% and 17\%. Assuming the radius of Io, the relative depth of the transit is $\sim (R_{\rm m}/R_{\rm s})^{2} \sim 5.7\times 10^{-4}$, which can be detected from dedicated space-borne photometry, given that the expected orbital period ranges between 5.5 and 12 hours.
Therefore, one or more planets orbiting LSR~J1835+3259 can be detected by a dedicated follow-up program \citep[cf.][]{Limbachetal21,Kottenetal26}. Another follow-up program could  adopt the technique described in \citet{Ozaetal24}, where monitoring Doppler shifts in the optical lines of Na or K,  combined with a search for transient activity, can be useful. Thus far, the Doppler shift technique has been confirmed by Keck  observations to indicate transient sodium clouds at $\sim$ $\pm 10-20$~km~s$^{-1}$ \citep{Unni2025}. Planetary systems have indeed been detected around UCD stars, notably in the case of TRAPPIST-1 \citep[cf.][]{Gillonetal16}. 

To keep a close-by planet volcanically active by tides, we need  forcing  its orbital eccentricity by an outer planet, so that a system of at least two planets is required. In principle, also the second planet could be directly detected, although its transit probability is lower owing to its larger distance from the star. However, its indirect detection could be feasible thanks to the transit timing variability induced in the inner planet. 

An alternative to the presence of a second planet could be an active moon that is losing mass and orbits  around a single planet inside the magnetosphere of LSR~J1835+3259. In this case, it would be the stellar gravitational perturbation that keeps the moon on an eccentric orbit around the planet; this would lead to heating by the tides induced by the planet \citep{Cassidyetal09}.
Giant or Neptune-mass planets orbiting low-mass stars are exceedingly rare \citep[e.g.,][]{Sabottaetal21,Pinamontietal22};  therefore, we considered a planet with the mass of the Earth orbited by a moon with a mass of $0.02$~M$_{\oplus}$. We analyze this system in Appendix~\ref{planet-moon} and find that the likely fate of the moon is to collide with its planet within a timescale much shorter than the age of the system, if we assume the planet to have a solid interior giving a modified tidal quality factor, $Q^{\prime}_{\rm m}$, of $ \sim 10^{3}$. On the other hand, if the planetary interior is fluid, $Q^{\prime}_{\rm m}$ can be greater by several orders of magnitude and the orbital decay timescale of the moon may become longer than the age of the system, thus making this a viable explanation. The detection of the planetary transits  should be feasible from the ground given that its expected radius is comparable with that of the planets orbiting TRAPPIST-1 or larger, but the detection of the small moon can be difficult,  although its presence can be directly signaled by the composition of its evaporating materials different from those of the planet. 

An interesting result that has a general relevance for UCD stars is that by \citet{KaoPineda25}, who show that the presence of a distant stellar or substellar companion is associated with an increase in radio detections in those stars. They suggested that such distant companions with typical separations on the order of $2-3$ au \citep[cf.][]{DupuyLiu17} could induce high-amplitude oscillations of the eccentricity of a close-by planet by the Lidov-Kozai mechanism when the orbit of the inner and outer binaries are sufficiently inclined or by the mechanism investigated by \citet{Mardling07} when they are coplanar. However, a necessary condition for those mechanisms to operate is that the period of the precession of the line of the apsides of the orbit of the body closer to the star (due to the stellar quadrupole moment and general relativity) be longer than the characteristic timescale of angular momentum exchange between the bodies in the system. Such a requirement is generally violated owing to the remarkable centrifugal distortion of  the star leading to an apsidal precession period  of a few tens or hundreds of years for the inner volcanically active planet, much shorter than the typical timescales of Kozai-Lidov cycles or angular momentum exchanges  in the Mardling mechanism that are on the order of $10^{4}$~yr. 

For this reason, we had to postulate the presence of another planet, much closer than any distant stellar companion, as the body exciting and maintaining the eccentric orbit of the tidally active innermost planet. However, the precession period of the line of the apsides of the outer planet is remarkably longer than that of the inner planet owing to its greater distance from the star. This may allow the Kozai mechanism to operate on the outer planet thus making its orbit eccentric as required by the Mardling model to account for the nonzero eccentricity of the inner planet. 

\section{Conclusions}

We adopted the ionospheric--magnetospheric coupling model by \citet{Turnpenneyetal17} and the Alfv\'en wave excitation model in the formulation of \citet{Noyolaetal14} to estimate the mass flow across the magnetosphere of the UCD star LSR~J1835+3259 that can account for its observed ECM emission. We considered the mass transport by corotating magnetospheric convection according to \citet{Hilletal81} to obtain an upper limit for the mass flow and considered several sources of plasma to the magnetosphere (see Table~\ref{table_comp} for a summary of the results). 

{For a closed magnetosphere, we find that a stellar ionospheric outflow on the order of $10^{5}$~kg~s$^{-1}$ can account for the ECM emission powered by ionospheric--magnetospheric coupling currents only if the effective ionospheric Pedersen conductance, $\Sigma_{\rm p}^{*}$, is lower than $\sim {0.003}$ mho, that appears to be outside the probable range for such a parameter as assumed by \citet{Turnpenneyetal17}}. Otherwise, a rocky planet with the radius and mass of the Jupiter moon Io, heated by internal tidal dissipation, appears to be a suitable source provided that  its orbital semimajor axis, $a$, is $ \la 6-14$ stellar radii. On the other hand, if the ECM emission is powered by the Alfv\'en waves excited by the orbital motion of such a putative planet, a  Pedersen conductance $\Sigma^{*}_{\rm p} \ga 0.1$~mho and $a \la (9-10) \, R_{\rm s}$ would lead to a mass supply rate of up to $\sim 10^{6}$~kg~s$^{-1}$. For smaller values of the Pedersen conductance, higher mass-loss rates would need to be supplied by the planet to account for the ECM emission (cf. Fig.~\ref{spi_density_fig1}). Its internal heating can be maintained only if an outer planet forces its orbital eccentricity, for example  according to the mechanism proposed by \citet{Mardling07}. Owing to its closeness to the star, the inner planet has a probability of $\sim 10-17$\% of transiting with an orbital period typically shorter than 12 hours. For a radius equal to that of Io, a transit depth of 500-600 parts per million is predicted; such a depth could be detected in a space-borne photometric follow-up, such as with  a telescope like CHEOPS \citep{Benzetal21}. Subtler  signatures could be visible by PLATO \citep{Raueretal25}. Moreover, such a  volcanically active planet could be indirectly detected by observing the absorption due to a circumstellar torus formed by its  evaporated matter, as in the case of Io \citep{Ozaetal24,Boehmetal25, Oza2026}. 

Close-by planetary systems orbiting UCD stars may explain why targets with a distant substellar or stellar companion have a remarkably higher probability of being detected as radio sources according to \citet{KaoPineda25}. Assuming the Kozai-Lidov mechanism and taking the distant companion into account could be crucial in establishing the eccentricity of the outer planet's  orbit, which in turn would establish  the eccentricity of the inner,  volcanically active planet.
  
The magnetospheric phenomena studied here provide complementary radio \citep{Narang_2023A} and visible-light \citep{Ozaetal19} signatures for the confirmation of putative satellites with radial velocity techniques, as inferred recently at the $\sim 37$~M$_{\rm J}$~BD CD-35~2722~B \citep{Hoyetal26}. Estimating the $P_{\rm tide}$ of the nearly  Jupiter-mass companion with a Jovian-like $Q^{\prime}_{\rm m} = 10^{5}$ and $R_{\rm m} \approx R_{\rm J}$ using Eq.~\eqref{tida_diss} yields $P_{\rm tide} \sim 10^{12}-10^{13}$~W ($\sim 0.01-0.1 \, P_{\rm Io}$) for the single eccentric companion solution of \citet{Hoyetal26} ($m \sin i \approx 0.9$~M$_{\rm J}$, $a \approx 0.20$~au, $e \approx 0.27$). Their alternative two-companion solution near a 2:1 mean-motion resonance yields $P_{\rm tide} \sim 10^{11}$~W ($\sim 10^{-3}\, P_{\rm Io}$) for the inner companion ($m \sin i \approx 0.22$~M$_{\rm J}$, $a \approx 0.125$~au, $e \approx 0.008$) and only $\sim 10^{8}$~W ($\sim 10^{-6}\, P_{\rm Io}$) for the nearly circular outer companion, far less than that of LSR~J1835+3259  (see Fig.~\ref{tidal_quant}). On the other hand, tidally heated close-in terrestrial satellites of UCDs and BDs may sustain volcanically fed magnetospheric plasma reservoirs. If so, they would be a promising probe of satellite systems beyond the Solar System. 
  
\begin{acknowledgements}

The authors are grateful to two anonymous referees, in particular to the second reviewer, for several comments and suggestions that helped to significantly improve their work. AVO expresses gratitude to G. Hallinan for insight on new observations and the asymmetric nature of LSR J1835+3259's synchrotron radio emission. We sincerely thank V. Dols for insight on electrodynamic processes at Jupiter-Io. AFL acknowledges support from the ASI-INAF agreement no.~2026-30-HH.0 “PLATO: preparation for the launch and phase E”.  
Research on exoplanets and stellar physics at Catania Astrophysical Observatory is funded by MUR (the Italian Ministry of University and Research) through the programs of INAF (the Italian National Institute for Astrophysics) including their grants for the support of fundamental research. 
The research described in this paper was carried out in part at the Jet Propulsion Laboratory, California Institute of Technology, under a contract with the National Aeronautics and Space Administration. © 2026. All rights reserved.

\end{acknowledgements}

\bibliographystyle{aa} 
\bibliography{refs} 

@ARTICLE{Rothetal24,
       author = {{Roth}, Lorenz and {Bl{\"o}cker}, Aljona and {de Kleer}, Katherine and {Goldstein}, David and {Lellouch}, Emmanuel and {Saur}, Joachim and {Schmidt}, Carl and {Strobel}, Darrell F. and {Tao}, Chihiro and {Tsuchiya}, Fuminori and {Dols}, Vincent and {Huybrighs}, Hans and {Mura}, Alessandro and {Szalay}, Jamey R. and {Badman}, Sarah V. and {de Pater}, Imke and {Dott}, Anne-Cathrine and {Kagitani}, Masato and {Klaiber}, Lea and {Koga}, Ryoichi and {McEwen}, Alfred S. and {Milby}, Zachariah and {Retherford}, Kurt D. and {Schlegel}, Stephan and {Thomas}, Nicolas and {Tseng}, Wei-Ling and {Vorburger}, Audrey},
        title = "{Mass Supply from Io to Jupiter's Magnetosphere}",
      journal = {\ssr},
         year = 2025,
        month = feb,
       volume = {221},
       number = {1},
          eid = {13},
        pages = {13},
          doi = {10.1007/s11214-025-01137-x},
archivePrefix = {arXiv},
       eprint = {2403.13970},
 primaryClass = {astro-ph.EP},
       adsurl = {https://ui.adsabs.harvard.edu/abs/2025SSRv..221...13R}
}

@ARTICLE{Kottenetal26,
       author = {{Kotten}, Brooke and {Limbach}, Mary Anne and {Vos}, Johanna M. and {Schrader}, Merle and {McCarthy}, Allison and {Kao}, Melodie M. and {Wilson}, Mikayla J. and {Householder}, Andrew and {Skemer}, Andrew and {Stevenson}, Kevin B. and {Su{\'a}rez}, Genaro and {Biller}, Beth and {Faherty}, Jacqueline and {Gonzales}, Eileen C. and {Muirhead}, Philip},
        title = "{On the Detectability of Volcanic Exo-Ios That May Fuel Auroras on Super-Jupiters}",
      journal = {\aj},
         year = 2026,
        month = sep,
       volume = {172},
       number = {3},
          eid = {167},
        pages = {167},
          doi = {10.3847/1538-3881/ae8adb},
archivePrefix = {arXiv},
       eprint = {2607.13030},
 primaryClass = {astro-ph.EP},
       adsurl = {https://ui.adsabs.harvard.edu/abs/2026AJ....172..167K}
}

@ARTICLE{Raueretal25,
       author = {{Rauer}, Heike and {Aerts}, Conny and {Cabrera}, Juan and {Deleuil}, Magali and {Erikson}, Anders and {Gizon}, Laurent and {Goupil}, Mariejo and {Heras}, Ana and {Walloschek}, Thomas and {Lorenzo-Alvarez}, Jose and {Marliani}, Filippo and {Martin-Garcia}, C{\'e}sar and {Mas-Hesse}, J. Miguel and {O'Rourke}, Laurence and {Osborn}, Hugh and {Pagano}, Isabella and {Piotto}, Giampaolo and {Pollacco}, Don and {Ragazzoni}, Roberto and {Ramsay}, Gavin and {Udry}, St{\'e}phane and {Appourchaux}, Thierry and {Benz}, Willy and {Brandeker}, Alexis and {G{\"u}del}, Manuel and {Janot-Pacheco}, Eduardo and {Kabath}, Petr and {Kjeldsen}, Hans and {Min}, Michiel and {Santos}, Nuno and {Smith}, Alan and {Suarez}, Juan-Carlos and {Werner}, Stephanie C. and {Aboudan}, Alessio and {Abreu}, Manuel and {Acu{\~n}a}, Lorena and {Adams}, Moritz and {Adibekyan}, Vardan and {Affer}, Laura and {Agneray}, Fran{\c{c}}ois and {Agnor}, Craig and {Aguirre B{\o}rsen-Koch}, Victor and {Ahmed}, Saad and {Aigrain}, Suzanne and {Al-Bahlawan}, Ashraf and {Alcacera Gil}, Ma de los Angeles and {Alei}, Eleonora and {Alencar}, Silvia and {Alexander}, Richard and {Alfonso-Garz{\'o}n}, Julia and {Alibert}, Yann and {Allende Prieto}, Carlos and {Almeida}, Leonardo and {Alonso Sobrino}, Roi and {Altavilla}, Giuseppe and {Althaus}, Christian and {Alvarez Trujillo}, Luis Alonso and {Amarsi}, Anish and {Ammler-von Eiff}, Matthias and {Am{\^o}res}, Eduardo and {Andrade}, Laerte and {Antoniadis-Karnavas}, Alexandros and {Ant{\'o}nio}, Carlos and {Aparicio del Moral}, Beatriz and {Appolloni}, Matteo and {Arena}, Claudio and {Armstrong}, David and {Aroca Aliaga}, Jose and {Asplund}, Martin and {Audenaert}, Jeroen and {Auricchio}, Natalia and {Avelino}, Pedro and {Baeke}, Ann and {Bailli{\'e}}, Kevin and {Balado}, Ana and {Ballber Balaguer{\'o}}, Pau and {Balestra}, Andrea and {Ball}, Warrick and {Ballans}, Herve and {Ballot}, Jerome and {Barban}, Caroline and {Barbary}, Ga{\"e}le and {Barbieri}, Mauro and {Barcel{\'o} Forteza}, Sebasti{\`a} and {Barker}, Adrian and {Barklem}, Paul and {Barnes}, Sydney and {Barrado Navascues}, David and {Barragan}, Oscar and {Baruteau}, Cl{\'e}ment and {Basu}, Sarbani and {Baudin}, Frederic and {Baumeister}, Philipp and {Bayliss}, Daniel and {Bazot}, Michael and {Beck}, Paul G. and {Belkacem}, Kevin and {Bellinger}, Earl and {Benatti}, Serena and {Benomar}, Othman and {B{\'e}rard}, Diane and {Bergemann}, Maria and {Bergomi}, Maria and {Bernardo}, Pierre and {Biazzo}, Katia and {Bignamini}, Andrea and {Bigot}, Lionel and {Billot}, Nicolas and {Binet}, Martin and {Biondi}, David and {Biondi}, Federico and {Birch}, Aaron C. and {Bitsch}, Bertram and {Bluhm Ceballos}, Paz Victoria and {B{\'o}di}, Attila and {Bogn{\'a}r}, Zs{\'o}fia and {Boisse}, Isabelle and {Bolmont}, Emeline and {Bonanno}, Alfio and {Bonavita}, Mariangela and {Bonfanti}, Andrea and {Bonfils}, Xavier and {Bonito}, Rosaria and {Bonomo}, Aldo Stefano and {B{\"o}rner}, Anko and {Boro Saikia}, Sudeshna and {Borreguero Mart{\'\i}n}, Elisa and {Borsa}, Francesco and {Borsato}, Luca and {Bossini}, Diego and {Bouchy}, Francois and {Bou{\'e}}, Gwena{\"e}l and {Boufleur}, Rodrigo and {Boumier}, Patrick and {Bourrier}, Vincent and {Bowman}, Dominic M. and {Bozzo}, Enrico and {Bradley}, Louisa and {Bray}, John and {Bressan}, Alessandro and {Breton}, Sylvain and {Brienza}, Daniele and {Brito}, Ana and {Brogi}, Matteo and {Brown}, Beverly and {Brown}, David J.~A. and {Brun}, Allan Sacha and {Bruno}, Giovanni and {Bruns}, Michael and {Buchhave}, Lars A. and {Bugnet}, Lisa and {Buldgen}, Ga{\"e}l and {Burgess}, Patrick and {Busatta}, Andrea and {Busso}, Giorgia and {Buzasi}, Derek and {Caballero}, Jos{\'e} A. and {Cabral}, Alexandre and {Cabrero Gomez}, Juan-Francisco and {Calderone}, Flavia and {Cameron}, Robert and {Cameron}, Andrew and {Campante}, Tiago and {Campos Gestal}, N{\'e}stor and {Canto Martins}, Bruno Leonardo and {Cara}, Christophe and {Carone}, Ludmila and {Carrasco}, Josep Manel and {Casagrande}, Luca and {Casewell}, Sarah L. and {Cassisi}, Santi and {Castellani}, Marco and {Castro}, Matthieu and {Catala}, Claude and {Catal{\'a}n Fern{\'a}ndez}, Irene and {Catelan}, M{\'a}rcio and {Cegla}, Heather and {Cerruti}, Chiara and {Cessa}, Virginie and {Chadid}, Merieme and {Chaplin}, William and {Charpinet}, Stephane and {Chiappini}, Cristina and {Chiarucci}, Simone and {Chiavassa}, Andrea and {Chinellato}, Simonetta and {Chirulli}, Giovanni and {Christensen-Dalsgaard}, J{\o}rgen and {Church}, Ross and {Claret}, Antonio and {Clarke}, Cathie and {Claudi}, Riccardo and {Clermont}, Lionel and {Coelho}, Hugo and {Coelho}, Joao and {Cogato}, Fabrizio and {Colom{\'e}}, Josep and {Condamin}, Mathieu and {Conde Garc{\'\i}a}, Fernando and {Conseil}, Simon},
        title = "{The PLATO mission}",
      journal = {Experimental Astronomy},
         year = 2025,
        month = jun,
       volume = {59},
       number = {3},
          eid = {26},
        pages = {26},
          doi = {10.1007/s10686-025-09985-9},
archivePrefix = {arXiv},
       eprint = {2406.05447},
 primaryClass = {astro-ph.IM},
       adsurl = {https://ui.adsabs.harvard.edu/abs/2025ExA....59...26R}
}

@ARTICLE{Hoyetal26,
       author = {{Hoy}, Kevin and {Zurlo}, Alice and {Pe{\~n}a R}, Pablo A. and {K{\"o}hler}, Jana and {Desidera}, Silvano and {Gratton}, Raffaele and {Lazzoni}, Cecilia and {Petrus}, Simon and {Rodler}, Florian and {Smoker}, Jonathan and {D'Orazi}, Valentina and {Carleo}, Ilaria and {Giovannini}, Ilaria},
        title = "{Planetary-mass exosatellite detected around the substellar companion of a star}",
      journal = {Nature},
         year = 2026,
        month = jul,
       volume = {655},
        pages = {865-869},
          doi = {10.1038/s41586-026-10751-w}
}

@article{McGrathJohnson89,
  author  = {McGrath, M. A. and Johnson, R. E.},
  title   = {Charge exchange cross sections for the Io plasma torus},
  journal = {J. Geophys. Res.}, volume = {94}, pages = {2677--2683}, year = {1989}}

@book{Johnson90,
  author    = {Johnson, R. E.},
  title     = {Energetic Charged-Particle Interactions with Atmospheres and Surfaces},
  series    = {Physics and Chemistry in Space}, volume = {19},
  publisher = {Springer-Verlag}, address = {Berlin}, year = {1990}}

@article{McGrathJohnson87,
  author  = {McGrath, M. A. and Johnson, R. E.},
  title   = {Magnetospheric plasma sputtering of Io's atmosphere},
  journal = {Icarus}, volume = {69}, pages = {519--531}, year = {1987}}

@ARTICLE{Dolsetal24,
       author = {{Dols}, V. and {Paterson}, W.~R. and {Bagenal}, F.},
        title = "{Parallel Electron Beams at Io: Numerical Simulations of the Dense Plasma Wake}",
      journal = {Journal of Geophysical Research (Space Physics)},
         year = 2024,
        month = mar,
       volume = {129},
       number = {3},
          eid = {e2023JA031763},
        pages = {e2023JA031763},
          doi = {10.1029/2023JA031763},
       adsurl = {https://ui.adsabs.harvard.edu/abs/2024JGRA..12931763D}
}

@INPROCEEDINGS{BagenalDols23,
       author = {{Bagenal}, Fran and {Dols}, Vincent},
        title = "{Space Environment of Io}",
    booktitle = {Io: A New View of Jupiter's Moon},
         year = 2023,
       editor = {{Lopes}, Rosaly M.~C. and {de Kleer}, Katherine and {Keane}, James Tuttle},
       series = {Astrophysics and Space Science Library},
       volume = {468},
        month = jan,
        pages = {291-322},
          doi = {10.1007/978-3-031-25670-7_9},
       adsurl = {https://ui.adsabs.harvard.edu/abs/2023ASSL..468..291B}
}

@ARTICLE{Sauretal99,
       author = {{Saur}, Joachim and {Neubauer}, Fritz M. and {Strobel}, Darrell F. and {Summers}, Michael E.},
        title = "{Three-dimensional plasma simulation of Io's interaction with the Io plasma torus: Asymmetric plasma flow}",
      journal = {\jgr},
         year = 1999,
        month = nov,
       volume = {104},
       number = {A11},
        pages = {25105-25126},
          doi = {10.1029/1999JA900304},
       adsurl = {https://ui.adsabs.harvard.edu/abs/1999JGR...10425105S}
}

@ARTICLE{Oza2026,
       author = {{Oza}, Apurva V. and {Gebek}, Andrea and {Meyer zu Westram}, Moritz and {Tokadjian}, Armen and {Piro}, Anthony L. and {Hu}, Renyu and {Unni}, Athira and {Chari}, Raghav and {Bello-Arufe}, Aaron and {Schmidt}, Carl A. and {Louca}, Amy J. and {Miguel}, Yamila and {Estrela}, Raissa and {Yang}, Jeehyun and {Damiano}, Mario and {Hasegawa}, Yasuhiro and {Welbanks}, Luis and {Powell}, Diana and {Garg}, Rishabh and {Gupta}, Pulkit and {Yung}, Yuk L. and {Lopes}, Rosaly M.~C.},
        title = "{Volcanic satellites tidally venting Na, K, and SO$_{2}$ in optical and infrared light}",
      journal = {\mnras},
         year = 2026,
        month = feb,
       volume = {546},
       number = {1},
          eid = {staf1526},
        pages = {staf1526},
          doi = {10.1093/mnras/staf1526},
archivePrefix = {arXiv},
       eprint = {2509.08349},
 primaryClass = {astro-ph.EP},
       adsurl = {https://ui.adsabs.harvard.edu/abs/2026MNRAS.546f1526O}
}

@ARTICLE{NicholsCowley04,
       author = {{Nichols}, J. and {Cowley}, S.},
        title = "{Magnetosphere-ionosphere coupling currents in Jupiter's middle magnetosphere: effect of precipitation-induced enhancement of the ionospheric Pedersen conductivity}",
      journal = {Annales Geophysicae},
         year = 2004,
        month = may,
       volume = {22},
       number = {5},
        pages = {1799-1827},
          doi = {10.5194/angeo-22-1799-2004},
       adsurl = {https://ui.adsabs.harvard.edu/abs/2004AnGeo..22.1799N}
}

@ARTICLE{Hallinanetal08,
       author = {{Hallinan}, G. and {Antonova}, A. and {Doyle}, J.~G. and {Bourke}, S. and {Lane}, C. and {Golden}, A.},
        title = "{Confirmation of the Electron Cyclotron Maser Instability as the Dominant Source of Radio Emission from Very Low Mass Stars and Brown Dwarfs}",
      journal = {\apj},
         year = 2008,
        month = sep,
       volume = {684},
       number = {1},
        pages = {644-653},
          doi = {10.1086/590360},
archivePrefix = {arXiv},
       eprint = {0805.4010},
 primaryClass = {astro-ph},
       adsurl = {https://ui.adsabs.harvard.edu/abs/2008ApJ...684..644H}
}

@ARTICLE{Bergeretal08a,
       author = {{Berger}, E. and {Gizis}, J.~E. and {Giampapa}, M.~S. and {Rutledge}, R.~E. and {Liebert}, J. and {Mart{\'\i}n}, E. and {Basri}, G. and {Fleming}, T.~A. and {Johns-Krull}, C.~M. and {Phan-Bao}, N. and et al.},
        title = "{Simultaneous Multiwavelength Observations of Magnetic Activity in Ultracool Dwarfs. I. The Complex Behavior of the M8.5 Dwarf TVLM 513-46546}",
      journal = {\apj},
         year = 2008,
        month = feb,
       volume = {673},
       number = {2},
        pages = {1080-1087},
          doi = {10.1086/524769},
archivePrefix = {arXiv},
       eprint = {0708.1511},
 primaryClass = {astro-ph},
       adsurl = {https://ui.adsabs.harvard.edu/abs/2008ApJ...673.1080B}
}

@ARTICLE{Metodievaetal17,
       author = {{Metodieva}, Y.~T. and {Kuznetsov}, A.~A. and {Antonova}, A.~E. and {Doyle}, J.~G. and {Ramsay}, G. and {Wu}, K.},
        title = "{Modelling the environment around five ultracool dwarfs via the radio domain}",
      journal = {\mnras},
         year = 2017,
        month = feb,
       volume = {465},
       number = {2},
        pages = {1995-2009},
          doi = {10.1093/mnras/stw2597},
archivePrefix = {arXiv},
       eprint = {1610.02989},
 primaryClass = {astro-ph.SR},
       adsurl = {https://ui.adsabs.harvard.edu/abs/2017MNRAS.465.1995M}
}

@ARTICLE{Scudderetal81,
       author = {{Scudder}, J.~D. and {Sittler}, E.~C. and {Bridge}, H.~S.},
        title = "{A survey of the plasma electron environment of Jupiter: a view from voyager}",
      journal = {\jgr},
         year = 1981,
        month = sep,
       volume = {86},
       number = {A10},
        pages = {8157-8179},
          doi = {10.1029/JA086iA10p08157},
       adsurl = {https://ui.adsabs.harvard.edu/abs/1981JGR....86.8157S}
}

@ARTICLE{Nichols11,
       author = {{Nichols}, J.~D.},
        title = "{Magnetosphere-ionosphere coupling at Jupiter-like exoplanets with internal plasma sources: implications for detectability of auroral radio emissions}",
      journal = {\mnras},
         year = 2011,
        month = jul,
       volume = {414},
       number = {3},
        pages = {2125-2138},
          doi = {10.1111/j.1365-2966.2011.18528.x},
archivePrefix = {arXiv},
       eprint = {1102.2737},
 primaryClass = {physics.space-ph},
       adsurl = {https://ui.adsabs.harvard.edu/abs/2011MNRAS.414.2125N}
}

@ARTICLE{MelroseDulk82,
       author = {{Melrose}, D.~B. and {Dulk}, G.~A.},
        title = "{Electron-cyclotron masers as the source of certain solar and stellar radio bursts.}",
      journal = {\apj},
         year = 1982,
        month = aug,
       volume = {259},
        pages = {844-858},
          doi = {10.1086/160219},
       adsurl = {https://ui.adsabs.harvard.edu/abs/1982ApJ...259..844M}
}

@ARTICLE{Vedanthametal20,
       author = {{Vedantham}, H.~K. and {Callingham}, J.~R. and {Shimwell}, T.~W. and {Dupuy}, T. and {Best}, William M.~J. and {Liu}, Michael C. and {Zhang}, Zhoujian and {De}, K. and {Lamy}, L. and {Zarka}, P. and et al.},
        title = "{Direct Radio Discovery of a Cold Brown Dwarf}",
      journal = {\apjl},
         year = 2020,
        month = nov,
       volume = {903},
       number = {2},
          eid = {L33},
        pages = {L33},
          doi = {10.3847/2041-8213/abc256},
archivePrefix = {arXiv},
       eprint = {2010.01915},
 primaryClass = {astro-ph.EP},
       adsurl = {https://ui.adsabs.harvard.edu/abs/2020ApJ...903L..33V}
}

@ARTICLE{Vedanthametal23,
       author = {{Vedantham}, H.~K. and {Dupuy}, T.~J. and {Evans}, E.~L. and {Sanghi}, A. and {Callingham}, J.~R. and {Shimwell}, T.~W. and {Best}, W.~M.~J. and {Liu}, M.~C. and {Zarka}, P.},
        title = "{Polarised radio pulsations from a new T-dwarf binary}",
      journal = {\aap},
         year = 2023,
        month = jul,
       volume = {675},
          eid = {L6},
        pages = {L6},
          doi = {10.1051/0004-6361/202244965},
archivePrefix = {arXiv},
       eprint = {2301.01003},
 primaryClass = {astro-ph.SR},
       adsurl = {https://ui.adsabs.harvard.edu/abs/2023A&A...675L...6V}
}

@ARTICLE{dePater81,
       author = {{de Pater}, I.},
        title = "{Radio maps of Jupiter's radiation belts and planetary disk at lambda 6 CM}",
      journal = {\aap},
         year = 1981,
        month = jan,
       volume = {93},
       number = {1-2},
        pages = {370-381},
       adsurl = {https://ui.adsabs.harvard.edu/abs/1981A&A....93..370D}
}

@ARTICLE{Zarka98,
       author = {{Zarka}, Philippe},
        title = "{Auroral radio emissions at the outer planets: Observations and theories}",
      journal = {\jgr},
         year = 1998,
        month = sep,
       volume = {103},
       number = {E9},
        pages = {20159-20194},
          doi = {10.1029/98JE01323},
       adsurl = {https://ui.adsabs.harvard.edu/abs/1998JGR...10320159Z}
}

@ARTICLE{Callinghmetal24,
       author = {{Callingham}, J.~R. and {Pope}, B.~J.~S. and {Kavanagh}, R.~D. and {Bellotti}, S. and {Daley-Yates}, S. and {Damasso}, M. and {Grie{\ss}meier}, J. -M. and {G{\"u}del}, M. and {G{\"u}nther}, M. and {Kao}, M.~M. and et al.},
        title = "{Radio signatures of star-planet interactions, exoplanets and space weather}",
      journal = {Nature Astronomy},
         year = 2024,
        month = nov,
       volume = {8},
        pages = {1359-1372},
          doi = {10.1038/s41550-024-02405-6},
archivePrefix = {arXiv},
       eprint = {2409.15507},
 primaryClass = {astro-ph.EP},
       adsurl = {https://ui.adsabs.harvard.edu/abs/2024NatAs...8.1359C}
}

@ARTICLE{Benzetal21,
       author = {{Benz}, W. and {Broeg}, C. and {Fortier}, A. and {Rando}, N. and {Beck}, T. and {Beck}, M. and {Queloz}, D. and {Ehrenreich}, D. and {Maxted}, P.~F.~L. and {Isaak}, K.~G. and et al.},
        title = "{The CHEOPS mission}",
      journal = {Experimental Astronomy},
         year = 2021,
        month = feb,
       volume = {51},
       number = {1},
        pages = {109-151},
          doi = {10.1007/s10686-020-09679-4},
archivePrefix = {arXiv},
       eprint = {2009.11633},
 primaryClass = {astro-ph.IM},
       adsurl = {https://ui.adsabs.harvard.edu/abs/2021ExA....51..109B}
}

@ARTICLE{Gillonetal16,
       author = {{Gillon}, Micha{\"e}l and {Jehin}, Emmanu{\"e}l and {Lederer}, Susan M. and {Delrez}, Laetitia and {de Wit}, Julien and {Burdanov}, Artem and {Van Grootel}, Val{\'e}rie and {Burgasser}, Adam J. and {Triaud}, Amaury H.~M.~J. and {Opitom}, Cyrielle and et al.},
        title = "{Temperate Earth-sized planets transiting a nearby ultracool dwarf star}",
      journal = {\nat},
         year = 2016,
        month = may,
       volume = {533},
       number = {7602},
        pages = {221-224},
          doi = {10.1038/nature17448},
archivePrefix = {arXiv},
       eprint = {1605.07211},
 primaryClass = {astro-ph.EP},
       adsurl = {https://ui.adsabs.harvard.edu/abs/2016Natur.533..221G}
}

@ARTICLE{Boehmetal25,
       author = {{Boehm}, V. Abby and {Seligman}, Darryl Z. and {Lewis}, Nikole K.},
        title = "{Constraining Ongoing Volcanic Outgassing Rates and Interior Compositions of Extrasolar Planets with Mass Measurements of Plasma Tori}",
      journal = {\apjl},
         year = 2025,
        month = jul,
       volume = {987},
       number = {1},
          eid = {L1},
        pages = {L1},
          doi = {10.3847/2041-8213/adde5e},
archivePrefix = {arXiv},
       eprint = {2506.08177},
 primaryClass = {astro-ph.EP},
       adsurl = {https://ui.adsabs.harvard.edu/abs/2025ApJ...987L...1B}
}

@ARTICLE{BagenalDols20,
       author = {{Bagenal}, Fran and {Dols}, Vincent},
        title = "{The Space Environment of Io and Europa}",
      journal = {Journal of Geophysical Research (Space Physics)},
         year = 2020,
        month = may,
       volume = {125},
       number = {5},
          eid = {e27485},
        pages = {e27485},
          doi = {10.1029/2019JA027485},
       adsurl = {https://ui.adsabs.harvard.edu/abs/2020JGRA..12527485B}
}

@ARTICLE{Okadaetal14,
       author = {{Okada}, Taku and {Iitaka}, Toshiaki and {Yagi}, Takehiko and {Aoki}, Katsutoshi},
        title = "{Electrical conductivity of ice VII}",
      journal = {Scientific Reports},
         year = 2014,
        month = jul,
       volume = {4},
          eid = {5778},
        pages = {5778},
          doi = {10.1038/srep05778},
       adsurl = {https://ui.adsabs.harvard.edu/abs/2014NatSR...4.5778O}
}

@ARTICLE{KaoPineda25,
       author = {{Kao}, Melodie M. and {Pineda}, J. Sebastian},
        title = "{Binarity enhances the occurrence rate of radiation belt emissions in ultracool dwarfs}",
      journal = {\mnras},
         year = 2025,
        month = may,
       volume = {539},
       number = {3},
        pages = {2292-2306},
          doi = {10.1093/mnras/stae905},
archivePrefix = {arXiv},
       eprint = {2403.08860},
 primaryClass = {astro-ph.SR},
       adsurl = {https://ui.adsabs.harvard.edu/abs/2025MNRAS.539.2292K}
}

@ARTICLE{SucerquiaCuello25f,
       author = {{Sucerquia}, Mario and {Cuello}, Nicol{\'a}s},
        title = "{Extreme exomoons in WASP-49 Ab: Dynamics and detectability}",
      journal = {\aap},
         year = 2025,
        month = feb,
       volume = {694},
          eid = {L8},
        pages = {L8},
          doi = {10.1051/0004-6361/202452968},
archivePrefix = {arXiv},
       eprint = {2501.05866},
 primaryClass = {astro-ph.EP},
       adsurl = {https://ui.adsabs.harvard.edu/abs/2025A&A...694L...8S}
}

@ARTICLE{Neubauer80,
       author = {{Neubauer}, F.~M.},
        title = "{Nonlinear standing Alfv{\'e}n wave current system at Io: Theory}",
      journal = {\jgr},
         year = 1980,
        month = mar,
       volume = {85},
       number = {A3},
        pages = {1171-1178},
          doi = {10.1029/JA085iA03p01171},
       adsurl = {https://ui.adsabs.harvard.edu/abs/1980JGR....85.1171N}
}

@ARTICLE{Soonetal00,
       author = {{Soon}, W. and {Baliunas}, S. and {Posmentier}, E.~S. and {Okeke}, P.},
        title = "{Variations of solar coronal hole area and terrestrial lower tropospheric air temperature from 1979 to mid-1998: astronomical forcings of change in earth's climate?}",
      journal = {\na},
         year = 2000,
        month = jan,
       volume = {4},
       number = {8},
        pages = {563-579},
          doi = {10.1016/S1384-1076(00)00002-6},
       adsurl = {https://ui.adsabs.harvard.edu/abs/2000NewA....4..563S}
}

@ARTICLE{Sauretal21,
       author = {{Saur}, Joachim and {Willmes}, Clarissa and {Fischer}, Christian and {Wennmacher}, Alexandre and {Roth}, Lorenz and {Youngblood}, Allison and {Strobel}, Darrell F. and {Reiners}, Ansgar},
        title = "{Brown dwarfs as ideal candidates for detecting UV aurora outside the Solar System: Hubble Space Telescope observations of 2MASS J1237+6526}",
      journal = {\aap},
         year = 2021,
        month = nov,
       volume = {655},
          eid = {A75},
        pages = {A75},
          doi = {10.1051/0004-6361/202040230},
archivePrefix = {arXiv},
       eprint = {2109.00827},
 primaryClass = {astro-ph.SR},
       adsurl = {https://ui.adsabs.harvard.edu/abs/2021A&A...655A..75S}
}

@ARTICLE{Neubauer98,
       author = {{Neubauer}, Fritz M.},
        title = "{The sub-Alfv{\'e}nic interaction of the Galilean satellites with the Jovian magnetosphere}",
      journal = {\jgr},
         year = 1998,
        month = sep,
       volume = {103},
       number = {E9},
        pages = {19843-19866},
          doi = {10.1029/97JE03370},
       adsurl = {https://ui.adsabs.harvard.edu/abs/1998JGR...10319843N}
}

@ARTICLE{FordRasio06,
       author = {{Ford}, Eric B. and {Rasio}, Frederic A.},
        title = "{On the Relation between Hot Jupiters and the Roche Limit}",
      journal = {\apjl},
         year = 2006,
        month = feb,
       volume = {638},
       number = {1},
        pages = {L45-L48},
          doi = {10.1086/500734},
archivePrefix = {arXiv},
       eprint = {astro-ph/0512632},
 primaryClass = {astro-ph},
       adsurl = {https://ui.adsabs.harvard.edu/abs/2006ApJ...638L..45F}
}

@ARTICLE{Zarkaetal04,
       author = {{Zarka}, P. and {Cecconi}, B. and {Kurth}, W.~S.},
        title = "{Jupiter's low-frequency radio spectrum from Cassini/Radio and Plasma Wave Science (RPWS) absolute flux density measurements}",
      journal = {Journal of Geophysical Research (Space Physics)},
         year = 2004,
        month = sep,
       volume = {109},
       number = {A9},
          eid = {A09S15},
        pages = {A09S15},
          doi = {10.1029/2003JA010260},
       adsurl = {https://ui.adsabs.harvard.edu/abs/2004JGRA..109.9S15Z}
}

@ARTICLE{QueinnecZarka01,
       author = {{Queinnec}, Julien and {Zarka}, Philippe},
        title = "{Flux, power, energy and polarization of Jovian S-bursts}",
      journal = {\planss},
         year = 2001,
        month = mar,
       volume = {49},
       number = {3-4},
        pages = {365-376},
          doi = {10.1016/S0032-0633(00)00157-4},
       adsurl = {https://ui.adsabs.harvard.edu/abs/2001P&SS...49..365Q}
}

@ARTICLE{Zarka07,
       author = {{Zarka}, Philippe},
        title = "{Plasma interactions of exoplanets with their parent star and associated radio emissions}",
      journal = {\planss},
         year = 2007,
        month = apr,
       volume = {55},
       number = {5},
        pages = {598-617},
          doi = {10.1016/j.pss.2006.05.045},
       adsurl = {https://ui.adsabs.harvard.edu/abs/2007P&SS...55..598Z}
}

@ARTICLE{Zarkaetal01,
       author = {{Zarka}, Philippe and {Treumann}, Rudolf A. and {Ryabov}, Boris P. and {Ryabov}, Vladimir B.},
        title = "{Magnetically-Driven Planetary Radio Emissions and Application to Extrasolar Planets}",
      journal = {\apss},
         year = 2001,
        month = jun,
       volume = {277},
        pages = {293-300},
          doi = {10.1023/A:1012221527425},
       adsurl = {https://ui.adsabs.harvard.edu/abs/2001Ap&SS.277..293Z}
}

@ARTICLE{Zhangetal23,
       author = {{Zhang}, Jiale and {Tian}, Hui and {Zarka}, Philippe and {Louis}, Corentin K. and {Lu}, Hongpeng and {Gao}, Dongyang and {Sun}, Xiaohui and {Yu}, Sijie and {Chen}, Bin and {Cheng}, Xin and {Wang}, Ke},
        title = "{Fine Structures of Radio Bursts from Flare Star AD Leo with FAST Observations}",
      journal = {\apj},
         year = 2023,
        month = aug,
       volume = {953},
       number = {1},
          eid = {65},
        pages = {65},
          doi = {10.3847/1538-4357/acdb77},
archivePrefix = {arXiv},
       eprint = {2306.00895},
 primaryClass = {astro-ph.SR},
       adsurl = {https://ui.adsabs.harvard.edu/abs/2023ApJ...953...65Z}
}

@ARTICLE{Hessetal07,
       author = {{Hess}, S. and {Zarka}, P. and {Mottez}, F.},
        title = "{Io Jupiter interaction, millisecond bursts and field-aligned potentials}",
      journal = {\planss},
         year = 2007,
        month = jan,
       volume = {55},
       number = {1-2},
        pages = {89-99},
          doi = {10.1016/j.pss.2006.05.016},
       adsurl = {https://ui.adsabs.harvard.edu/abs/2007P&SS...55...89H}
}

@ARTICLE{Kuznetsovetal12,
       author = {{Kuznetsov}, A.~A. and {Doyle}, J.~G. and {Yu}, S. and {Hallinan}, G. and {Antonova}, A. and {Golden}, A.},
        title = "{Comparative Analysis of Two Formation Scenarios of Bursty Radio Emission from Ultracool Dwarfs}",
      journal = {\apj},
         year = 2012,
        month = feb,
       volume = {746},
       number = {1},
          eid = {99},
        pages = {99},
          doi = {10.1088/0004-637X/746/1/99},
archivePrefix = {arXiv},
       eprint = {1111.7019},
 primaryClass = {astro-ph.SR},
       adsurl = {https://ui.adsabs.harvard.edu/abs/2012ApJ...746...99K}
}

@ARTICLE{Letoetal17,
       author = {{Leto}, P. and {Trigilio}, C. and {Buemi}, C.~S. and {Umana}, G. and {Ingallinera}, A. and {Cerrigone}, L.},
        title = "{Probing the magnetosphere of the M8.5 dwarf TVLM 513-46546 by modelling its auroral radio emission. Hint of star exoplanet interaction?}",
      journal = {\mnras},
         year = 2017,
        month = aug,
       volume = {469},
       number = {2},
        pages = {1949-1967},
          doi = {10.1093/mnras/stx995},
archivePrefix = {arXiv},
       eprint = {1611.00511},
 primaryClass = {astro-ph.SR},
       adsurl = {https://ui.adsabs.harvard.edu/abs/2017MNRAS.469.1949L}
}

@ARTICLE{Hill01,
       author = {{Hill}, T.~W.},
        title = "{The Jovian auroral oval}",
      journal = {\jgr},
         year = 2001,
        month = may,
       volume = {106},
       number = {A5},
        pages = {8101-8108},
          doi = {10.1029/2000JA000302},
       adsurl = {https://ui.adsabs.harvard.edu/abs/2001JGR...106.8101H}
}

@ARTICLE{Schrijver09,
       author = {{Schrijver}, Carolus J.},
        title = "{On a Transition from Solar-Like Coronae to Rotation-Dominated Jovian-Like Magnetospheres in Ultracool Main-Sequence Stars}",
      journal = {\apjl},
         year = 2009,
        month = jul,
       volume = {699},
       number = {2},
        pages = {L148-L152},
          doi = {10.1088/0004-637X/699/2/L148},
archivePrefix = {arXiv},
       eprint = {0905.1354},
 primaryClass = {astro-ph.SR},
       adsurl = {https://ui.adsabs.harvard.edu/abs/2009ApJ...699L.148S}
}

@ARTICLE{Bergeretal08,
       author = {{Berger}, E. and {Basri}, G. and {Gizis}, J.~E. and {Giampapa}, M.~S. and {Rutledge}, R.~E. and {Liebert}, J. and {Mart{\'\i}n}, E. and {Fleming}, T.~A. and {Johns-Krull}, C.~M. and {Phan-Bao}, N. and {Sherry}, W.~H.},
        title = "{Simultaneous Multiwavelength Observations of Magnetic Activity in Ultracool Dwarfs. II. Mixed Trends in VB 10 and LSR 1835+32 and the Possible Role of Rotation}",
      journal = {\apj},
         year = 2008,
        month = apr,
       volume = {676},
       number = {2},
        pages = {1307-1318},
          doi = {10.1086/529131},
archivePrefix = {arXiv},
       eprint = {0710.3383},
 primaryClass = {astro-ph},
       adsurl = {https://ui.adsabs.harvard.edu/abs/2008ApJ...676.1307B}
}

@ARTICLE{Turnpenneyetal17,
       author = {{Turnpenney}, S. and {Nichols}, J.~D. and {Wynn}, G.~A. and {Casewell}, S.~L.},
        title = "{Auroral radio emission from ultracool dwarfs: a Jovian model}",
      journal = {\mnras},
         year = 2017,
        month = oct,
       volume = {470},
       number = {4},
        pages = {4274-4284},
          doi = {10.1093/mnras/stx1508},
archivePrefix = {arXiv},
       eprint = {1706.04679},
 primaryClass = {astro-ph.SR},
       adsurl = {https://ui.adsabs.harvard.edu/abs/2017MNRAS.470.4274T}
}

@ARTICLE{Vidotto21,
       author = {{Vidotto}, Aline A.},
        title = "{The evolution of the solar wind}",
      journal = {Living Reviews in Solar Physics},
         year = 2021,
        month = dec,
       volume = {18},
       number = {1},
          eid = {3},
        pages = {3},
          doi = {10.1007/s41116-021-00029-w},
archivePrefix = {arXiv},
       eprint = {2103.15748},
 primaryClass = {astro-ph.SR},
       adsurl = {https://ui.adsabs.harvard.edu/abs/2021LRSP...18....3V}
}

@ARTICLE{Hilletal81,
       author = {{Hill}, T.~W. and {Dessler}, A.~J. and {Maher}, L.~J.},
        title = "{Corotating magnetospheric convection}",
      journal = {\jgr},
         year = 1981,
        month = oct,
       volume = {86},
       number = {A11},
        pages = {9020-9028},
          doi = {10.1029/JA086iA11p09020},
       adsurl = {https://ui.adsabs.harvard.edu/abs/1981JGR....86.9020H}
}

@ARTICLE{SiscoeSummers81,
       author = {{Siscoe}, G.~L. and {Summers}, D.},
        title = "{Centrifugally driven diffusion of Iogenic plasma}",
      journal = {\jgr},
         year = 1981,
        month = sep,
       volume = {86},
       number = {A10},
        pages = {8471-8479},
          doi = {10.1029/JA086iA10p08471},
       adsurl = {https://ui.adsabs.harvard.edu/abs/1981JGR....86.8471S}
}

@ARTICLE{Schmidt22,
       author = {{Schmidt}, Carl A.},
        title = "{Doppler-Shifted Alkali D Absorption as Indirect Evidence for Exomoons}",
      journal = {Frontiers in Astronomy and Space Sciences},
         year = 2022,
        month = mar,
       volume = {9},
          eid = {801873},
        pages = {801873},
          doi = {10.3389/fspas.2022.801873},
archivePrefix = {arXiv},
       eprint = {2202.13815},
 primaryClass = {astro-ph.EP},
       adsurl = {https://ui.adsabs.harvard.edu/abs/2022FrASS...901873S}
}

@ARTICLE{Haffetal81,
       author = {{Haff}, P.~K. and {Watson}, C.~C. and {Yung}, Y.~L.},
        title = "{Sputter ejection of matter from Io}",
      journal = {\jgr},
         year = 1981,
        month = aug,
       volume = {86},
       number = {A8},
        pages = {6933-6938},
          doi = {10.1029/JA086iA08p06933},
       adsurl = {https://ui.adsabs.harvard.edu/abs/1981JGR....86.6933H}
}

@ARTICLE{Johnson04,
       author = {{Johnson}, R.~E.},
        title = "{The Magnetospheric Plasma-driven Evolution of Satellite Atmospheres}",
      journal = {\apjl},
         year = 2004,
        month = jul,
       volume = {609},
       number = {2},
        pages = {L99-L102},
          doi = {10.1086/422912},
       adsurl = {https://ui.adsabs.harvard.edu/abs/2004ApJ...609L..99J}
}

@ARTICLE{GoldreichNicholson77,
       author = {{Goldreich}, P. and {Nicholson}, P.~D.},
        title = "{Turbulent Viscosity and Jupiter's Tidal Q}",
      journal = {\icarus},
         year = 1977,
        month = feb,
       volume = {30},
       number = {2},
        pages = {301-304},
          doi = {10.1016/0019-1035(77)90163-4},
       adsurl = {https://ui.adsabs.harvard.edu/abs/1977Icar...30..301G}
}

@ARTICLE{Wu05,
       author = {{Wu}, Yanqin},
        title = "{Origin of Tidal Dissipation in Jupiter. II. The Value of Q}",
      journal = {\apj},
         year = 2005,
        month = dec,
       volume = {635},
       number = {1},
        pages = {688-710},
          doi = {10.1086/497355},
archivePrefix = {arXiv},
       eprint = {astro-ph/0407628},
 primaryClass = {astro-ph},
       adsurl = {https://ui.adsabs.harvard.edu/abs/2005ApJ...635..688W}
}

@ARTICLE{Berdyuginaetal17,
       author = {{Berdyugina}, S.~V. and {Harrington}, D.~M. and {Kuzmychov}, O. and {Kuhn}, J.~R. and {Hallinan}, G. and {Kowalski}, A.~F. and {Hawley}, S.~L.},
        title = "{First Detection of a Strong Magnetic Field on a Bursty Brown Dwarf: Puzzle Solved}",
      journal = {\apj},
         year = 2017,
        month = sep,
       volume = {847},
       number = {1},
          eid = {61},
        pages = {61},
          doi = {10.3847/1538-4357/aa866b},
archivePrefix = {arXiv},
       eprint = {1709.02861},
 primaryClass = {astro-ph.SR},
       adsurl = {https://ui.adsabs.harvard.edu/abs/2017ApJ...847...61B}
}

@ARTICLE{Kaoetal23,
       author = {{Kao}, Melodie M. and {Mioduszewski}, Amy J. and {Villadsen}, Jackie and {Shkolnik}, Evgenya L.},
        title = "{Resolved imaging confirms a radiation belt around an ultracool dwarf}",
      journal = {\nat},
         year = 2023,
        month = jul,
       volume = {619},
       number = {7969},
        pages = {272-275},
          doi = {10.1038/s41586-023-06138-w},
archivePrefix = {arXiv},
       eprint = {2302.12841},
 primaryClass = {astro-ph.EP},
       adsurl = {https://ui.adsabs.harvard.edu/abs/2023Natur.619..272K}
}

@ARTICLE{Noyolaetal14,
       author = {{Noyola}, J.~P. and {Satyal}, S. and {Musielak}, Z.~E.},
        title = "{Detection of Exomoons through Observation of Radio Emissions}",
      journal = {\apj},
         year = 2014,
        month = aug,
       volume = {791},
       number = {1},
          eid = {25},
        pages = {25},
          doi = {10.1088/0004-637X/791/1/25},
archivePrefix = {arXiv},
       eprint = {1308.4184},
 primaryClass = {astro-ph.EP},
       adsurl = {https://ui.adsabs.harvard.edu/abs/2014ApJ...791...25N}
}

@ARTICLE{Sauretal13,
       author = {{Saur}, J. and {Grambusch}, T. and {Duling}, S. and {Neubauer}, F.~M. and {Simon}, S.},
        title = "{Magnetic energy fluxes in sub-Alfv{\'e}nic planet star and moon planet interactions}",
      journal = {\aap},
         year = 2013,
        month = apr,
       volume = {552},
          eid = {A119},
        pages = {A119},
          doi = {10.1051/0004-6361/201118179},
       adsurl = {https://ui.adsabs.harvard.edu/abs/2013A&A...552A.119S}
}

@ARTICLE{Kollmanetal18,
       author = {{Kollmann}, P. and {Roussos}, E. and {Paranicas}, C. and {Woodfield}, E.~E. and {Mauk}, B.~H. and {Clark}, G. and {Smith}, D.~C. and {Vandegriff}, J.},
        title = "{Electron Acceleration to MeV Energies at Jupiter and Saturn}",
      journal = {Journal of Geophysical Research (Space Physics)},
         year = 2018,
        month = nov,
       volume = {123},
       number = {11},
        pages = {9110-9129},
          doi = {10.1029/2018JA025665},
       adsurl = {https://ui.adsabs.harvard.edu/abs/2018JGRA..123.9110K}
}

@ARTICLE{Climentetal23,
       author = {{Climent}, J.~B. and {Guirado}, J.~C. and {P{\'e}rez-Torres}, M. and {Marcaide}, J.~M. and {Pe{\~n}a-Mo{\~n}ino}, L.},
        title = "{Evidence for a radiation belt around a brown dwarf}",
      journal = {Science},
         year = 2023,
        month = sep,
       volume = {381},
       number = {6662},
        pages = {1120-1124},
          doi = {10.1126/science.adg6635},
archivePrefix = {arXiv},
       eprint = {2303.06453},
 primaryClass = {astro-ph.SR},
       adsurl = {https://ui.adsabs.harvard.edu/abs/2023Sci...381.1120C}
}

@ARTICLE{dePaterDunn03,
       author = {{de Pater}, Imke and {Dunn}, David E.},
        title = "{VLA observations of Jupiter's synchrotron radiation at 15 and 22 GHz}",
      journal = {\icarus},
         year = 2003,
        month = jun,
       volume = {163},
       number = {2},
        pages = {449-455},
          doi = {10.1016/S0019-1035(03)00068-X},
       adsurl = {https://ui.adsabs.harvard.edu/abs/2003Icar..163..449D}
}

@ARTICLE{Milleretal09,
       author = {{Miller}, N. and {Fortney}, J.~J. and {Jackson}, B.},
        title = "{Inflating and Deflating Hot Jupiters: Coupled Tidal and Thermal Evolution of Known Transiting Planets}",
      journal = {\apj},
         year = 2009,
        month = sep,
       volume = {702},
       number = {2},
        pages = {1413-1427},
          doi = {10.1088/0004-637X/702/2/1413},
archivePrefix = {arXiv},
       eprint = {0907.1268},
 primaryClass = {astro-ph.EP},
       adsurl = {https://ui.adsabs.harvard.edu/abs/2009ApJ...702.1413M}
}

@ARTICLE{Lanza21,
       author = {{Lanza}, A.~F.},
        title = "{An internal heating mechanism operating in ultra-short-period planets orbiting magnetically active stars}",
      journal = {\aap},
         year = 2021,
        month = sep,
       volume = {653},
          eid = {A112},
        pages = {A112},
          doi = {10.1051/0004-6361/202140284},
archivePrefix = {arXiv},
       eprint = {2107.03044},
 primaryClass = {astro-ph.EP},
       adsurl = {https://ui.adsabs.harvard.edu/abs/2021A&A...653A.112L}
}

@ARTICLE{Kislyakovaetal17,
       author = {{Kislyakova}, K.~G. and {Noack}, L. and {Johnstone}, C.~P. and {Zaitsev}, V.~V. and {Fossati}, L. and {Lammer}, H. and {Khodachenko}, M.~L. and {Odert}, P. and {G{\"u}del}, M.},
        title = "{Magma oceans and enhanced volcanism on TRAPPIST-1 planets due to induction heating}",
      journal = {Nature Astronomy},
         year = 2017,
        month = oct,
       volume = {1},
        pages = {878-885},
          doi = {10.1038/s41550-017-0284-0},
archivePrefix = {arXiv},
       eprint = {1710.08761},
 primaryClass = {astro-ph.EP},
       adsurl = {https://ui.adsabs.harvard.edu/abs/2017NatAs...1..878K}
}

@ARTICLE{YoshinoKatsura13,
       author = {{Yoshino}, Takashi and {Katsura}, Tomoo},
        title = "{Electrical Conductivity of Mantle Minerals: Role of Water in Conductivity Anomalies}",
      journal = {Annual Review of Earth and Planetary Sciences},
         year = 2013,
        month = may,
       volume = {41},
        pages = {605-628},
          doi = {10.1146/annurev-earth-050212-124022},
       adsurl = {https://ui.adsabs.harvard.edu/abs/2013AREPS..41..605Y}
}

@ARTICLE{Route17,
       author = {{Route}, Matthew},
        title = "{Radio-flaring Ultracool Dwarf Population Synthesis}",
      journal = {\apj},
         year = 2017,
        month = aug,
       volume = {845},
       number = {1},
          eid = {66},
        pages = {66},
          doi = {10.3847/1538-4357/aa7ede},
archivePrefix = {arXiv},
       eprint = {1707.02212},
 primaryClass = {astro-ph.SR},
       adsurl = {https://ui.adsabs.harvard.edu/abs/2017ApJ...845...66R}
}

@ARTICLE{KaoShkolnik24,
       author = {{Kao}, Melodie M. and {Shkolnik}, Evgenya L.},
        title = "{The occurrence rate of quiescent radio emission for ultracool dwarfs using a generalized semi-analytical Bayesian framework}",
      journal = {\mnras},
         year = 2024,
        month = jan,
       volume = {527},
       number = {3},
        pages = {6835-6866},
          doi = {10.1093/mnras/stad2272},
archivePrefix = {arXiv},
       eprint = {2306.16460},
 primaryClass = {astro-ph.SR},
       adsurl = {https://ui.adsabs.harvard.edu/abs/2024MNRAS.527.6835K}
}

@ARTICLE{Kaoetal19,
       author = {{Kao}, Melodie M. and {Hallinan}, Gregg and {Pineda}, J. Sebastian},
        title = "{Constraints on magnetospheric radio emission from Y dwarfs}",
      journal = {\mnras},
         year = 2019,
        month = aug,
       volume = {487},
       number = {2},
        pages = {1994-2004},
          doi = {10.1093/mnras/stz1372},
       adsurl = {https://ui.adsabs.harvard.edu/abs/2019MNRAS.487.1994K}
}

@ARTICLE{Kaoetal18,
       author = {{Kao}, Melodie M. and {Hallinan}, Gregg and {Pineda}, J. Sebastian and {Stevenson}, David and {Burgasser}, Adam},
        title = "{The Strongest Magnetic Fields on the Coolest Brown Dwarfs}",
      journal = {\apjs},
         year = 2018,
        month = aug,
       volume = {237},
       number = {2},
          eid = {25},
        pages = {25},
          doi = {10.3847/1538-4365/aac2d5},
archivePrefix = {arXiv},
       eprint = {1808.02485},
 primaryClass = {astro-ph.SR},
       adsurl = {https://ui.adsabs.harvard.edu/abs/2018ApJS..237...25K}
}

@ARTICLE{Kaoetal16,
       author = {{Kao}, Melodie M. and {Hallinan}, Gregg and {Pineda}, J. Sebastian and {Escala}, Ivanna and {Burgasser}, Adam and {Bourke}, Stephen and {Stevenson}, David},
        title = "{Auroral Radio Emission from Late L and T Dwarfs: A New Constraint on Dynamo Theory in the Substellar Regime}",
      journal = {\apj},
         year = 2016,
        month = feb,
       volume = {818},
       number = {1},
          eid = {24},
        pages = {24},
          doi = {10.3847/0004-637X/818/1/24},
archivePrefix = {arXiv},
       eprint = {1511.03661},
 primaryClass = {astro-ph.SR},
       adsurl = {https://ui.adsabs.harvard.edu/abs/2016ApJ...818...24K}
}

@ARTICLE{Tyleretal15,
       author = {{Tyler}, Robert H. and {Henning}, Wade G. and {Hamilton}, Christopher W.},
        title = "{Tidal Heating in a Magma Ocean within Jupiter's Moon Io}",
      journal = {\apjs},
         year = 2015,
        month = jun,
       volume = {218},
       number = {2},
          eid = {22},
        pages = {22},
          doi = {10.1088/0067-0049/218/2/22},
       adsurl = {https://ui.adsabs.harvard.edu/abs/2015ApJS..218...22T}
}

@ARTICLE{Unni2025,
       author = {{Unni}, Athira and {Oza}, Apurva V. and {Hoeijmakers}, H. Jens and {Seidel}, Julia V. and {Sivarani}, Thirupathi and {Schmidt}, Carl A. and {Kesseli}, Aurora Y. and {de Kleer}, Katherine and {Baker}, Ashley D. and {Gebek}, Andrea and {Westram}, Moritz Meyer zu and {Fisher}, Chloe and {Sallum}, Steph and {Bestha}, Manjunath and {Bello-Arufe}, Aaron},
        title = "{Doppler shifted transient sodium detection by KECK/HIRES}",
      journal = {\mnras},
         year = 2025,
        month = jun,
       volume = {540},
       number = {1},
        pages = {L48-L53},
          doi = {10.1093/mnrasl/slaf031},
archivePrefix = {arXiv},
       eprint = {2504.03974},
 primaryClass = {astro-ph.EP},
       adsurl = {https://ui.adsabs.harvard.edu/abs/2025MNRAS.540L..48U}
}

@ARTICLE{kisare2024,
       author = {{Kisare}, A.~M. and {Fabrycky}, D.~C.},
        title = "{Tidal dissipation in satellites prevents Hill sphere escape}",
      journal = {\mnras},
         year = 2024,
        month = jan,
       volume = {527},
       number = {3},
        pages = {4371-4377},
          doi = {10.1093/mnras/stad3543},
archivePrefix = {arXiv},
       eprint = {2309.11609},
 primaryClass = {astro-ph.EP},
       adsurl = {https://ui.adsabs.harvard.edu/abs/2024MNRAS.527.4371K}
}

@ARTICLE{Narang_2023A,
       author = {{Narang}, Mayank and {Oza}, Apurva V. and {Hakim}, Kaustubh and {Manoj}, P. and {Banyal}, Ravinder K. and {Thorngren}, Daniel P.},
        title = "{Radio-loud Exoplanet-exomoon Survey: GMRT Search for Electron Cyclotron Maser Emission}",
      journal = {\aj},
         year = 2023,
        month = jan,
       volume = {165},
       number = {1},
          eid = {1},
        pages = {1},
          doi = {10.3847/1538-3881/ac9eb8},
archivePrefix = {arXiv},
       eprint = {2210.13298},
 primaryClass = {astro-ph.EP},
       adsurl = {https://ui.adsabs.harvard.edu/abs/2023AJ....165....1N}
}

@ARTICLE{Narangetal23,
       author = {{Narang}, Mayank and {Oza}, Apurva V. and {Hakim}, Kaustubh and {Manoj}, P. and {Tyagi}, Himanshu and {Banerjee}, Bihan and {Surya}, Arun and {Nayak}, Prasanta K. and {Banyal}, Ravinder K. and {Thorngren}, Daniel P.},
        title = "{uGMRT observations of the hot-Saturn WASP-69b: Radio-Loud Exoplanet-Exomoon Survey II (RLEES II)}",
      journal = {\mnras},
         year = 2023,
        month = jun,
       volume = {522},
       number = {2},
        pages = {1662-1668},
          doi = {10.1093/mnras/stad1027},
archivePrefix = {arXiv},
       eprint = {2303.17269},
 primaryClass = {astro-ph.EP},
       adsurl = {https://ui.adsabs.harvard.edu/abs/2023MNRAS.522.1662N}
}

@ARTICLE{Ozaetal19,
       author = {{Oza}, Apurva V. and {Johnson}, Robert E. and {Lellouch}, Emmanuel and {Schmidt}, Carl and {Schneider}, Nick and {Huang}, Chenliang and {Gamborino}, Diana and {Gebek}, Andrea and {Wyttenbach}, Aurelien and {Demory}, Brice-Olivier and {Mordasini}, Christoph and {Saxena}, Prabal and {Dubois}, David and {Moullet}, Arielle and {Thomas}, Nicolas},
        title = "{Sodium and Potassium Signatures of Volcanic Satellites Orbiting Close-in Gas Giant Exoplanets}",
      journal = {\apj},
         year = 2019,
        month = nov,
       volume = {885},
       number = {2},
          eid = {168},
        pages = {168},
          doi = {10.3847/1538-4357/ab40cc},
archivePrefix = {arXiv},
       eprint = {1908.10732},
 primaryClass = {astro-ph.EP},
       adsurl = {https://ui.adsabs.harvard.edu/abs/2019ApJ...885..168O}
}

@ARTICLE{Gudel04,
       author = {{G{\"u}del}, Manuel},
        title = "{X-ray astronomy of stellar coronae}",
      journal = {\aapr},
         year = 2004,
        month = sep,
       volume = {12},
       number = {2-3},
        pages = {71-237},
          doi = {10.1007/s00159-004-0023-2},
archivePrefix = {arXiv},
       eprint = {astro-ph/0406661},
 primaryClass = {astro-ph},
       adsurl = {https://ui.adsabs.harvard.edu/abs/2004A&ARv..12...71G}
}

@ARTICLE{Limbachetal21,
       author = {{Limbach}, Mary Anne and {Vos}, Johanna M. and {Winn}, Joshua N. and {Heller}, Ren{\'e} and {Mason}, Jeffrey C. and {Schneider}, Adam C. and {Dai}, Fei},
        title = "{On the Detection of Exomoons Transiting Isolated Planetary-mass Objects}",
      journal = {\apjl},
         year = 2021,
        month = sep,
       volume = {918},
       number = {2},
          eid = {L25},
        pages = {L25},
          doi = {10.3847/2041-8213/ac1e2d},
archivePrefix = {arXiv},
       eprint = {2108.08323},
 primaryClass = {astro-ph.EP},
       adsurl = {https://ui.adsabs.harvard.edu/abs/2021ApJ...918L..25L}
}

@ARTICLE{Jacksonetal08,
       author = {{Jackson}, Brian and {Greenberg}, Richard and {Barnes}, Rory},
        title = "{Tidal Evolution of Close-in Extrasolar Planets}",
      journal = {\apj},
         year = 2008,
        month = may,
       volume = {678},
       number = {2},
        pages = {1396-1406},
          doi = {10.1086/529187},
archivePrefix = {arXiv},
       eprint = {0802.1543},
 primaryClass = {astro-ph},
       adsurl = {https://ui.adsabs.harvard.edu/abs/2008ApJ...678.1396J}
}

@ARTICLE{Mardling07,
       author = {{Mardling}, Rosemary A.},
        title = "{Long-term tidal evolution of short-period planets with companions}",
      journal = {\mnras},
         year = 2007,
        month = dec,
       volume = {382},
       number = {4},
        pages = {1768-1790},
          doi = {10.1111/j.1365-2966.2007.12500.x},
archivePrefix = {arXiv},
       eprint = {0706.0224},
 primaryClass = {astro-ph},
       adsurl = {https://ui.adsabs.harvard.edu/abs/2007MNRAS.382.1768M}
}

@ARTICLE{MardlingLin02,
       author = {{Mardling}, Rosemary A. and {Lin}, D.~N.~C.},
        title = "{Calculating the Tidal, Spin, and Dynamical Evolution of Extrasolar Planetary Systems}",
      journal = {\apj},
         year = 2002,
        month = jul,
       volume = {573},
       number = {2},
        pages = {829-844},
          doi = {10.1086/340752},
       adsurl = {https://ui.adsabs.harvard.edu/abs/2002ApJ...573..829M}
}

@ARTICLE{DupuyLiu17,
       author = {{Dupuy}, Trent J. and {Liu}, Michael C.},
        title = "{Individual Dynamical Masses of Ultracool Dwarfs}",
      journal = {\apjs},
         year = 2017,
        month = aug,
       volume = {231},
       number = {2},
          eid = {15},
        pages = {15},
          doi = {10.3847/1538-4365/aa5e4c},
archivePrefix = {arXiv},
       eprint = {1703.05775},
 primaryClass = {astro-ph.SR},
       adsurl = {https://ui.adsabs.harvard.edu/abs/2017ApJS..231...15D}
}

@ARTICLE{Cillibrasietal21,
       author = {{Cilibrasi}, M. and {Szul{\'a}gyi}, J. and {Grimm}, S.~L. and {Mayer}, L.},
        title = "{An N-body population synthesis framework for the formation of moons around Jupiter-like planets}",
      journal = {\mnras},
         year = 2021,
        month = jul,
       volume = {504},
       number = {4},
        pages = {5455-5474},
          doi = {10.1093/mnras/stab1179},
archivePrefix = {arXiv},
       eprint = {2011.11513},
 primaryClass = {astro-ph.EP},
       adsurl = {https://ui.adsabs.harvard.edu/abs/2021MNRAS.504.5455C}
}

@ARTICLE{Magauddaetal24,
       author = {{Magaudda}, E. and {Stelzer}, B. and {Osten}, R.~A. and {Pineda}, J.~S. and {Raetz}, St. and {McKay}, M.},
        title = "{Transitions in magnetic behavior at the substellar boundary}",
      journal = {\aap},
         year = 2024,
        month = jul,
       volume = {687},
          eid = {A95},
        pages = {A95},
          doi = {10.1051/0004-6361/202449403},
archivePrefix = {arXiv},
       eprint = {2401.17292},
 primaryClass = {astro-ph.SR},
       adsurl = {https://ui.adsabs.harvard.edu/abs/2024A&A...687A..95M}
}

@ARTICLE{GuedelBenz93,
       author = {{Guedel}, Manuel and {Benz}, Arnold O.},
        title = "{X-Ray/Microwave Relation of Different Types of Active Stars}",
      journal = {\apjl},
         year = 1993,
        month = mar,
       volume = {405},
        pages = {L63},
          doi = {10.1086/186766},
       adsurl = {https://ui.adsabs.harvard.edu/abs/1993ApJ...405L..63G}
}

@ARTICLE{Saur03,
       author = {{Saur}, Joachim and {Strobel}, Darrell F. and {Neubauer}, Fritz M. and {Summers}, Michael E.},
        title = "{The ion mass loading rate at Io}",
      journal = {\icarus},
         year = 2003,
        month = jun,
       volume = {163},
       number = {2},
        pages = {456-468},
          doi = {10.1016/S0019-1035(03)00085-X},
       adsurl = {https://ui.adsabs.harvard.edu/abs/2003Icar..163..456S}
}

@ARTICLE{Jacksonetal08a,
       author = {{Jackson}, Brian and {Greenberg}, Richard and {Barnes}, Rory},
        title = "{Tidal Heating of Extrasolar Planets}",
      journal = {\apj},
         year = 2008,
        month = jul,
       volume = {681},
       number = {2},
        pages = {1631-1638},
          doi = {10.1086/587641},
archivePrefix = {arXiv},
       eprint = {0803.0026},
 primaryClass = {astro-ph},
       adsurl = {https://ui.adsabs.harvard.edu/abs/2008ApJ...681.1631J}
}

@ARTICLE{Laineyetal09,
       author = {{Lainey}, Val{\'e}ry and {Arlot}, Jean-Eudes and {Karatekin}, {\"O}zg{\"u}r and {van Hoolst}, Tim},
        title = "{Strong tidal dissipation in Io and Jupiter from astrometric observations}",
      journal = {\nat},
         year = 2009,
        month = jun,
       volume = {459},
       number = {7249},
        pages = {957-959},
          doi = {10.1038/nature08108},
       adsurl = {https://ui.adsabs.harvard.edu/abs/2009Natur.459..957L}
}

@ARTICLE{Veederetal94,
       author = {{Veeder}, Glenn J. and {Matson}, Dennis L. and {Johnson}, Torrence V. and {Blaney}, Diana L. and {Goguen}, Jay D.},
        title = "{Io's heat flow from infrared radiometry: 1983-1993}",
      journal = {\jgr},
         year = 1994,
        month = aug,
       volume = {99},
       number = {E8},
        pages = {17095-17162},
          doi = {10.1029/94JE00637},
       adsurl = {https://ui.adsabs.harvard.edu/abs/1994JGR....9917095V}
}

@ARTICLE{Domingosetal06,
       author = {{Domingos}, R.~C. and {Winter}, O.~C. and {Yokoyama}, T.},
        title = "{Stable satellites around extrasolar giant planets}",
      journal = {\mnras},
         year = 2006,
        month = dec,
       volume = {373},
       number = {3},
        pages = {1227-1234},
          doi = {10.1111/j.1365-2966.2006.11104.x},
       adsurl = {https://ui.adsabs.harvard.edu/abs/2006MNRAS.373.1227D}
}

@ARTICLE{Cassidyetal09,
       author = {{Cassidy}, Timothy A. and {Mendez}, Rolando and {Arras}, Phil and {Johnson}, Robert E. and {Skrutskie}, Michael F.},
        title = "{Massive Satellites of Close-In Gas Giant Exoplanets}",
      journal = {\apj},
         year = 2009,
        month = oct,
       volume = {704},
       number = {2},
        pages = {1341-1348},
          doi = {10.1088/0004-637X/704/2/1341},
archivePrefix = {arXiv},
       eprint = {0909.0770},
 primaryClass = {astro-ph.EP},
       adsurl = {https://ui.adsabs.harvard.edu/abs/2009ApJ...704.1341C}
}

@ARTICLE{Sucerquiaetal20,
       author = {{Sucerquia}, Mario and {Ram{\'\i}rez}, Vanesa and {Alvarado-Montes}, Jaime A. and {Zuluaga}, Jorge I.},
        title = "{Can close-in giant exoplanets preserve detectable moons?}",
      journal = {\mnras},
         year = 2020,
        month = mar,
       volume = {492},
       number = {3},
        pages = {3499-3508},
          doi = {10.1093/mnras/stz3548},
archivePrefix = {arXiv},
       eprint = {1912.08049},
 primaryClass = {astro-ph.EP},
       adsurl = {https://ui.adsabs.harvard.edu/abs/2020MNRAS.492.3499S}
}

@ARTICLE{Alvarado-Montesetal17,
       author = {{Alvarado-Montes}, J.~A. and {Zuluaga}, Jorge I. and {Sucerquia}, Mario},
        title = "{The effect of close-in giant planets' evolution on tidal-induced migration of exomoons}",
      journal = {\mnras},
         year = 2017,
        month = nov,
       volume = {471},
       number = {3},
        pages = {3019-3027},
          doi = {10.1093/mnras/stx1745},
archivePrefix = {arXiv},
       eprint = {1707.02906},
 primaryClass = {astro-ph.EP},
       adsurl = {https://ui.adsabs.harvard.edu/abs/2017MNRAS.471.3019A}
}

@ARTICLE{Sabottaetal21,
       author = {{Sabotta}, S. and {Schlecker}, M. and {Chaturvedi}, P. and {Guenther}, E.~W. and {Mu{\~n}oz Rodr{\'\i}guez}, I. and {Mu{\~n}oz S{\'a}nchez}, J.~C. and {Caballero}, J.~A. and {Shan}, Y. and {Reffert}, S. and {Ribas}, I. and {Reiners}, A. and {Hatzes}, A.~P. and {Amado}, P.~J. and {Klahr}, H. and {Morales}, J.~C. and {Quirrenbach}, A. and {Henning}, Th. and {Dreizler}, S. and {Pall{\'e}}, E. and {Perger}, M. and {Azzaro}, M. and {Jeffers}, S.~V. and {Kaminski}, A. and {K{\"u}rster}, M. and {Lafarga}, M. and {Montes}, D. and {Passegger}, V.~M. and {Zechmeister}, M.},
        title = "{The CARMENES search for exoplanets around M dwarfs. Planet occurrence rates from a subsample of 71 stars}",
      journal = {\aap},
         year = 2021,
        month = sep,
       volume = {653},
          eid = {A114},
        pages = {A114},
          doi = {10.1051/0004-6361/202140968},
archivePrefix = {arXiv},
       eprint = {2107.03802},
 primaryClass = {astro-ph.EP},
       adsurl = {https://ui.adsabs.harvard.edu/abs/2021A&A...653A.114S}
}

@ARTICLE{Pinamontietal22,
       author = {{Pinamonti}, M. and {Sozzetti}, A. and {Maldonado}, J. and {Affer}, L. and {Micela}, G. and {Bonomo}, A.~S. and {Lanza}, A.~F. and {Perger}, M. and {Ribas}, I. and {Gonz{\'a}lez Hern{\'a}ndez}, J.~I. and {Bignamini}, A. and {Claudi}, R. and {Covino}, E. and {Damasso}, M. and {Desidera}, S. and {Giacobbe}, P. and {Gonz{\'a}lez-{\'A}lvarez}, E. and {Herrero}, E. and {Leto}, G. and {Maggio}, A. and {Molinari}, E. and {Morales}, J.~C. and {Pagano}, I. and {Petralia}, A. and {Piotto}, G. and {Poretti}, E. and {Rebolo}, R. and {Scandariato}, G. and {Su{\'a}rez Mascare{\~n}o}, A. and {Toledo-Padr{\'o}n}, B. and {Zanmar S{\'a}nchez}, R.},
        title = "{HADES RV Programme with HARPS-N at TNG. XV. Planetary occurrence rates around early-M dwarfs}",
      journal = {\aap},
         year = 2022,
        month = aug,
       volume = {664},
          eid = {A65},
        pages = {A65},
          doi = {10.1051/0004-6361/202142828},
archivePrefix = {arXiv},
       eprint = {2203.04648},
 primaryClass = {astro-ph.EP},
       adsurl = {https://ui.adsabs.harvard.edu/abs/2022A&A...664A..65P}
}

@ARTICLE{CollierCameronJardine18,
       author = {{Collier Cameron}, Andrew and {Jardine}, Moira},
        title = "{Hierarchical Bayesian calibration of tidal orbit decay rates among hot Jupiters}",
      journal = {\mnras},
         year = 2018,
        month = may,
       volume = {476},
       number = {2},
        pages = {2542-2555},
          doi = {10.1093/mnras/sty292},
archivePrefix = {arXiv},
       eprint = {1801.10561},
 primaryClass = {astro-ph.EP},
       adsurl = {https://ui.adsabs.harvard.edu/abs/2018MNRAS.476.2542C}
}

@ARTICLE{Bigg64,
       author = {{Bigg}, E.~K.},
        title = "{Influence of the Satellite Io on Jupiter's Decametric Emission}",
      journal = {\nat},
         year = 1964,
        month = sep,
       volume = {203},
       number = {4949},
        pages = {1008-1010},
          doi = {10.1038/2031008a0},
       adsurl = {https://ui.adsabs.harvard.edu/abs/1964Natur.203.1008B}
}

@ARTICLE{BrownChaffee74,
       author = {{Brown}, Robert A. and {Chaffee}, Jr., Frederic H.},
        title = "{High-Resolution Spectra of Sodium Emission from IO}",
      journal = {\apjl},
         year = 1974,
        month = feb,
       volume = {187},
        pages = {L125},
          doi = {10.1086/181413},
       adsurl = {https://ui.adsabs.harvard.edu/abs/1974ApJ...187L.125B}
}

@ARTICLE{Morabitoetal79,
       author = {{Morabito}, L.~A. and {Synnott}, S.~P. and {Kupferman}, P.~N. and {Collins}, S.~A.},
        title = "{Discovery of Currently Active Extraterrestrial Volcanism}",
      journal = {Science},
         year = 1979,
        month = jun,
       volume = {204},
       number = {4396},
        pages = {972},
          doi = {10.1126/science.204.4396.972},
       adsurl = {https://ui.adsabs.harvard.edu/abs/1979Sci...204..972M}
}

@ARTICLE{Ozaetal24,
       author = {{Oza}, Apurva V. and {Seidel}, Julia V. and {Hoeijmakers}, H. Jens and {Unni}, Athira and {Kesseli}, Aurora Y. and {Schmidt}, Carl A. and {Sivarani}, Thirupathi and {Bello-Arufe}, Aaron and {Gebek}, Andrea and {Meyer zu Westram}, Moritz and {Sousa}, S{\'e}rgio G. and {Lopes}, Rosaly M.~C. and {Hu}, Renyu and {de Kleer}, Katherine and {Fisher}, Chloe and {Charnoz}, S{\'e}bastien and {Baker}, Ashley D. and {Halverson}, Samuel P. and {Schneider}, Nick M. and {Psaridi}, Angelica and {Wyttenbach}, Aur{\'e}lien and {Torres}, Santiago and {Bhatnagar}, Ishita and {Johnson}, Robert E.},
        title = "{Redshifted Sodium Transient near Exoplanet Transit}",
      journal = {\apjl},
         year = 2024,
        month = oct,
       volume = {973},
       number = {2},
          eid = {L53},
        pages = {L53},
          doi = {10.3847/2041-8213/ad6b29},
archivePrefix = {arXiv},
       eprint = {2409.19844},
 primaryClass = {astro-ph.EP},
       adsurl = {https://ui.adsabs.harvard.edu/abs/2024ApJ...973L..53O}
}

@ARTICLE{Parketal24,
       author = {{Park}, R.~S. and {Jacobson}, R.~A. and {Gomez Casajus}, L. and {Nimmo}, F. and {Ermakov}, A.~I. and {Keane}, J.~T. and {McKinnon}, W.~B. and {Stevenson}, D.~J. and {Akiba}, R. and {Idini}, B. and {Buccino}, D.~R. and {Magnanini}, A. and {Parisi}, M. and {Tortora}, P. and {Zannoni}, M. and {Mura}, A. and {Durante}, D. and {Iess}, L. and {Connerney}, J.~E.~P. and {Levin}, S.~M. and {Bolton}, S.~J.},
        title = "{Io's tidal response precludes a shallow magma ocean}",
      journal = {\nat},
         year = 2025,
        month = feb,
       volume = {638},
       number = {8049},
        pages = {69-73},
          doi = {10.1038/s41586-024-08442-5},
       adsurl = {https://ui.adsabs.harvard.edu/abs/2025Natur.638...69P}
}

@ARTICLE{GebekOza20,
       author = {{Gebek}, Andrea and {Oza}, Apurva V.},
        title = "{Alkaline exospheres of exoplanet systems: evaporative transmission spectra}",
      journal = {\mnras},
         year = 2020,
        month = oct,
       volume = {497},
       number = {4},
        pages = {5271-5291},
          doi = {10.1093/mnras/staa2193},
archivePrefix = {arXiv},
       eprint = {2005.02536},
 primaryClass = {astro-ph.EP},
       adsurl = {https://ui.adsabs.harvard.edu/abs/2020MNRAS.497.5271G}
}

@ARTICLE{Schmidtetal23,
       author = {{Schmidt}, Carl and {Sharov}, Mikhail and {de Kleer}, Katherine and {Schneider}, Nick and {de Pater}, Imke and {Phipps}, Phillip H. and {Conrad}, Albert and {Moore}, Luke and {Withers}, Paul and {Spencer}, John and {Morgenthaler}, Jeff and {Ilyin}, Ilya and {Strassmeier}, Klaus and {Veillet}, Christian and {Hill}, John and {Brown}, Mike},
        title = "{Io's Optical Aurorae in Jupiter's Shadow}",
      journal = {Planetary Science Journal},
         year = 2023,
        month = feb,
       volume = {4},
       number = {2},
          eid = {36},
        pages = {36},
          doi = {10.3847/PSJ/ac85b0},
archivePrefix = {arXiv},
       eprint = {2302.10849},
 primaryClass = {astro-ph.EP},
       adsurl = {https://ui.adsabs.harvard.edu/abs/2023PSJ.....4...36S}
}

\begin{appendix}

\section{Magnetospheric mass flux due to an ionospheric mass outflow}
\label{stellar_wind}

The X-ray emission in UCD stars is regarded to come from hot coronal plasma located close to their surfaces,  according to \citet{Schrijver09} and \citet{Magauddaetal24}. Specifically, the expected X-ray luminosity can be estimated as \citep[cf. Eq. 3 in][]{Schrijver09}
\begin{equation}
L_{\rm X} \sim 4\pi R_{\rm s}^{2} H_{\rm p} n_{\rm e0}^{2} \Lambda(T_{\rm cor}),
\label{corona_lum}
\end{equation}
where $R_{\rm s}$ is the radius of the star, $H_{\rm p}$ the coronal pressure scale height, $n_{\rm e0}$ the electron number density at the base of the corona, and $\Lambda(T_{\rm cor})$ the radiative loss function at the coronal temperature $T_{\rm cor}$. The high thermal conductivity in the closed corona or along open magnetic flux tubes connecting it with the ionosphere and the magnetosphere,  makes the plasma approximately isothermal along each magnetic field line close to the star. As a typical coronal temperature, we assumed  $T_{\rm cor} =1.5$~MK to evaluate $\Lambda (T_{\rm cor})$ \citep{Schrijver09,Vidotto21}. The pressure scale height is $H_{\rm p} = \tilde{R}T_{\rm cor}/(\tilde{\mu} g)$, where $\tilde{R}$ is the gas constant, $\tilde{\mu} =0.6$, the mean molecular weight assumed for a solar composition, and $g=GM_{\rm s}/R_{\rm s}^{2}$ the acceleration of gravity at the surface of the star. 

In a reference frame rotating with the angular velocity ${\vec \Omega}_{\rm s}$ of the star,  the momentum equation for a stationary ionospheric outflow is 
\begin{equation}
\rho ({\vec \varv} \cdot \nabla) {\vec \varv} = -\nabla p + {\vec J} \times {\vec B} +\rho \nabla \Psi - 2\rho \, {\vec \Omega}_{\rm s} \times {\vec \varv},  
\label{mom_eq_wind}
\end{equation}
where $\rho$ is the plasma density, $\vec \varv$ the flow velocity, $p$ the plasma pressure, $\vec J$ the current density, $\vec B$ the magnetic field, and $\Psi$ the total potential (gravitational plus centrifugal). Close to the star, the stellar magnetic field is so strong that the outflow is everywhere directed along the magnetic field lines that are fixed in the rotating frame. Therefore, ${\vec \varv} = \varv \hat{\vec s}$ and ${\vec B} = B \hat{\vec s}$, where $\hat{\vec s}$ is the unit vector pointing in the direction of the magnetic field at any given point. 

The projection of Eq.~\eqref{mom_eq_wind} along the unit vector $\hat{\vec s}$ gives 
\begin{equation}
    \frac{1}{2} \frac{\partial \varv ^{2}}{\partial s} = -\frac{1}{\rho}\frac{\partial p}{\partial s} +  \frac{\partial \Psi}{\partial s},
\label{mom_line}
\end{equation}
where we made use of the identity $\nabla (\varv^{2}/2) =  ({\vec \varv} \cdot \nabla) {\vec \varv} + {\vec \varv} \times (\nabla \times {\vec \varv)} $. 
The  plasma follows the ideal gas law $p = \varv_{\rm c}^{2} \rho$, where 
\begin{equation}
  \varv_{\rm c}^{2} \equiv (\tilde{R} T_{\rm w}/\tilde{\mu}) = 8.3 \times 10^{3} \left(\frac{T_{\rm w}}{\tilde{\mu}} \right)\, \mbox{m$^{2}$~s$^{-2}$}
  \label{sound_speed}
\end{equation}
is the square of the isothermal sound speed that is a constant in our isothermal  outflow model, where $T_{\rm w} \sim T_{\rm cor}$ is the  outflow plasma temperature.  

By substituting the ideal gas law into Eq.~\eqref{mom_line}, we find
\begin{equation}
\frac{\partial}{\partial s} \left( \frac{1}{2}\varv^{2} + \varv^{2}_{\rm c}\ln \rho - \Psi \right) = 0.  
\label{bernoulli}
\end{equation}
This can be immediately integrated along a given field line to give the density of the outflow:   
\begin{equation}
    \rho = \rho_{0} \exp \left( \frac{\Psi-\Psi_{0}}{\varv_{\rm c}^{2}} \right) \exp{\left[-\frac{1}{2}\left(\frac{\varv}{\varv_{\rm c}}\right)^{2}\right]},
    \label{density_eq}
\end{equation}
where $\rho_{0}$ is the density at the base of the field line where the potential is $\Psi_{0}$ and $\varv= 0$. The value of $\rho_{0}$ (or an upper limit for it) can be obtained from the X-ray luminosity (or an upper limit to the X-ray luminosity) for a given coronal temperature from Eq.~\eqref{corona_lum} by multiplying the number density $n_{\rm e0}$ by the mean atomic mass of the plasma. 
The total potential is given by
\begin{equation}
    \Psi = \frac{GM_{\rm s}}{r} + \frac{1}{2} \Omega^{2} r^{2} \sin^{2} \theta,
    \label{potential}
\end{equation}
where $r$ is the distance from the center of the star, and $\theta$ the colatitude measured from the north pole, which we assumed (for the sake of simplicity) to coincide with the pole of the magnetic field dipole of the star. The potential at the base of the field lines on the stellar surface can be assumed to be uniform and equal to $\Psi_{0} = GM_{\rm s}/R_{\rm s}$, provided that the centrifugal deformation of the star $\epsilon \equiv \Omega_{\rm s}^{2} R_{\rm s}^{3}/(2GM_{\rm s}) \ll 1$.  The corotation radius $R_{\rm cor}$ as given by Eq.~\eqref{rcorot} in the equatorial plane corresponds to the minimum of the total potential  and hence of the density along the radial coordinate as can be seen from the vanishing of the first derivative of Eq.~\eqref{potential}.

We assumed that the outflow can be described by Eq.~\eqref{bernoulli} up to the corotation radius, $R_{\rm cor}$, that is, in the domain where the gravity is stronger than the centrifugal force and the denser plasma tends to move toward the star. Conversely, in the domain beyond the corotation radius, i.e., for $r \ge R_{\rm cor}$,  magnetospheric convection prevails producing an outward mass flux, fed by the ionospheric outflow coming from the domain below. In the domain where $r \geq R_{\rm cor}$, we applied the Hill model assuming a constant mass flux across the magnetosphere that is ruled by the ionospheric outflow because it determines the boundary conditions applied at the inner boundary of the domain at $r=R_{\rm cor}$. Such an assumption is an approximation because it does not consider the outflow along the field lines whose  footpoints are at high latitudes and cross the equatorial plane at radial distances greater than $R_{\rm cor}$. Rigorously speaking, the consequence of such a high-latitude contribution to the outflow is that the $\dot{M}$ term in Eqs.~(16) and~(17) of \citet{Turnpenneyetal17} is not independent of the radial distance as required by the Hill model. Nevertheless, for the sake of simplicity, we continued to apply the Hill model assuming that the mass flow across the magnetosphere, $\dot{M}$, can be regarded as approximately constant for $r\geq R_{\rm cor}$. The centrifugal force becomes aligned with the magnetic field of a dipole at a latitude of $35\fdg5$ thus effectively accelerating the outflow that maps to an equatorial distance of 1.5~$R_{\rm s}$. On the other hand, the footpoint of the field line mapping on the equatorial plane at the corotation radius $R_{\rm cor}=3.87~R_{\rm s}$ in the case of our star is at a latitude of $59\fdg4$. Therefore, a significant fraction of the mass outflow is indeed coming from closer than $R_{\rm cor} = 3.87\,R_{\rm s}$ in the case of our star, although, rigorously speaking, the contribution of the outflow along the field lines whose footpoints are at latitudes $\geq 59\fdg4$  cannot be neglected since it is comparable with that originating from the lower latitudes.

With the caveat mentioned above and applying the Hill picture, the total mass flux across the magnetosphere originated by the ionospheric outflow can be expressed as 
\begin{eqnarray}
    \dot{M}_{\rm ion\, flow} & \sim & f_{\rm w} \int_{S(R_{\rm cor})} \rho(R_{\rm cor}) \varv_{\rm mc}(R_{\rm cor}) \, dS \sim  \nonumber \\
    & \sim & 4\pi f_{\rm w}\, R_{\rm cor} H_{\rm c} \, \rho(R_{\rm cor}) \varv_{\rm mc}(R_{\rm cor}),
    \label{mass_loss_rate_wind}
\end{eqnarray}
where $0 \leq f_{\rm w} \leq 1$ is the fraction of the coronal base covered by the open magnetic field lines along which the outflow develops,  $S(R_{\rm cor})$ the cylindrical surface of radius $R_{\rm cor}$ and height $2H_{\rm c}$ across which the mass flow occurs, $\rho (R_{\rm cor})$ the outflow density at the corotation radius that we estimated using Eqs.~\eqref{density_eq} and~\eqref{potential} with $r=R_{\rm cor}$ because the above outflow model can be applied for $r \leq R_{\rm cor}$. The exponential factor containing the outflow velocity $\varv$ in Eq.~\eqref{density_eq} can be taken to be unity because we are interested in an upper limit for the mass-loss rate and $\varv \ga \varv_{\rm c}$ at $r=R_{\rm cor}$ on the equatorial plane. 

In the case of our outflow, the centrifugal scale height is evaluated by means of Eq.~\eqref{cen_hp} where the temperature $T$ is the outflow temperature ($T=T_{\rm w}$), while the particle mass is that of the proton because the stellar corona, from which the flow originates, is mostly composed of hydrogen ions. 
By combining Eqs.~\eqref{v_mconv}, \eqref{r_hill}, and~\eqref{mass_loss_rate_wind}, we obtain an explicit expression for the mass loss flux across the magnetosphere supplied by the stellar ionospheric outflow $\dot{M}_{\rm ion \, flow}$
as
\begin{equation}
\dot{M}_{\rm ion\, flow} \sim  \frac{4\pi f_{\rm w}^{2}}{3} \left( \frac{R_{\rm cor}}{R_{\rm s}} \right)^{6} \frac{\left[ R_{\rm cor} H_{\rm c} \, \Omega_{\rm s} \, \rho(R_{\rm cor}) \right]^{2}}{\Sigma_{\rm p}^{*} B_{\rm s}^{2}}. 
\label{wind_mass_loss_fin}
\end{equation}
Given our neglect of the density decrease produced by the rightmost exponential factor in Eq.~\eqref{density_eq} and the upper limit to the magnetospheric convection velocity as given by  Eq.~\eqref{v_mconv}, Eq.~\eqref{wind_mass_loss_fin} provides an upper limit to the mass flow across the magnetosphere that can be supplied by a stellar ionospheric flow.

\section{Induction heating}
\label{induction_model}

Given the strong magnetic fields observed in radio-emitting UCDs and BDs, we postulate that an additional internal heating source for our putative planet may be  electromagnetic induction due to a time-varying magnetic field across its section  during the rotation of the star and its orbital motion. 
A detailed model of the electromagnetic induction and the associated heating inside a planetary body has been presented by, for example, \citet{Kislyakovaetal17}. The stellar magnetic field as seen by the planetary interior is oscillating with a frequency $\omega = | 2\pi/P_{\rm rot} - 2\pi/P_{\rm orb}|$, where $P_{\rm rot}$ is the rotation period of the star, while $P_{\rm orb}$ is the orbital period of our  planet. The time-varying stellar field is attenuated inside the orbiting planet  because it decreases exponentially as $\exp(-d/\delta)$, where $d$ is the depth from the surface and $\delta$ is the electromagnetic skin depth. In the regime of interest for our problem, the skin depth is given by 
\begin{equation}
    \delta = \min \left(\sqrt{\frac{2}{\mu \sigma \omega}}, R_{\rm m} \right),  
\end{equation} 
where $\sigma$ is the interior electric conductivity of the planet. 

\begin{figure}
\hspace{0.75cm}
\centerline{
\includegraphics[width=11cm,height=8cm]{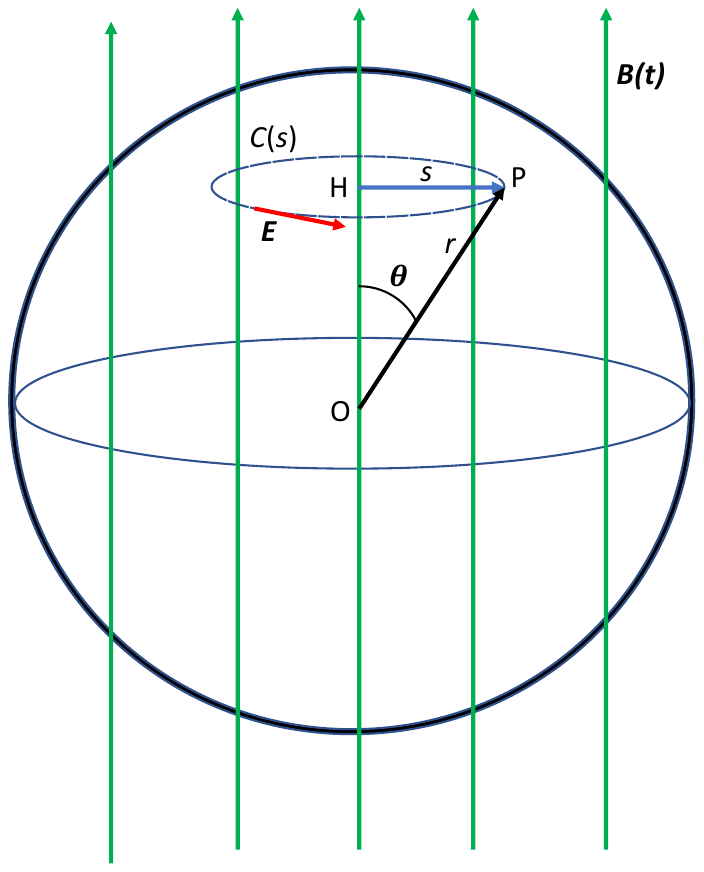}}
\caption{Sketch of the electric induction inside a planet with a uniform internal magnetic field equal to the outer uniform field produced by the star the planet is orbiting, assuming that the skin depth is remarkably greater than the radius of the planet.  In the reference frame of the planet, the external field, $\vec B(t)$, depends on the time, $t$. The local induced electric field, $\vec E$, is  tangent to a circle $C(s)$ (dashed line) lying in a  plane orthogonal to the magnetic field, $\vec B$, and with radius $s= r \sin \theta$, where $r$ is the radial distance from the center of the planet, $O$, and $\theta$  the colatitude of a generic point,  $P$, along $C(s)$. Thanks to the cylindrical symmetry of our system around the direction of the magnetic field and its uniformity, the induction law applied to the circle $C(s)$ gives $2\pi s E = - \pi s^{2} \dot{B}(t)$, from which Eq.~\eqref{E_field} follows.}
\label{figure1}
\end{figure}

The magnetic field inside the planet can be computed if we know the radial profile of its electric conductivity according to the model by \citet{Kislyakovaetal17}. Nevertheless, given our ignorance about its interior composition and being interested only in an order of magnitude estimate of the internal heating produced by the electromagnetic induction, we used a much simpler approach, assuming that the internal magnetic field is equal to the external stellar field down to a depth $\delta$ inside the planet. 
We sketch the physical system in Figure~\ref{figure1}. Assuming the same  uniform magnetic field inside and outside a spherical planet and given the symmetry of the system, the electric field, $\vec E$, at a distance $s = r \sin \theta$ from the polar axis, which we assumed to be parallel to the magnetic field, is given by the induction law as 
\begin{equation}
E = - \frac{1}{2}\, \dot{B}(t)\,r \sin \theta ,
\label{E_field}
\end{equation}
where $r$ is the distance from the center of the planet, $\theta$ the colatitude, and $\dot{B}(t) \equiv dB/dt$ the time derivative of the magnetic field inside the planet.
According to the Ohm law, the current density $J = \sigma E$, where $\sigma$ is the local electric conductivity of the planet that we assumed to depend only on the radial coordinate, $r$. Therefore, the  Joule heating per elementary volume, $dV$, is given by 
\begin{equation}
dP_{\rm diss} = \frac{J^{2}}{\sigma} dV = \sigma E^{2}\ dV = \frac{1}{4} (\dot{B})^{2}\, \sigma(r) \, r^{2} \sin^{2} \theta \,  dV,
\end{equation}
where the volume element for  given $r$ and $\theta$ is $dV = 2\pi r^{2} \sin \theta \, d\theta \, dr$ thanks to the cylindrical symmetry of the system that allows us to integrate immediately over the azimuthal angle. 
By integrating over the colatitude, we obtained the power dissipated inside the elementary spherical shell between $r$ and $r+dr$ as
\begin{equation}
    dP_{\rm diss} = \frac{2\pi}{3} \sigma (r) \, (\dot{B})^{2} \,r^{4}  dr.
    \label{dPdr}
\end{equation}
Adopting a sinusoidal time variation for the magnetic field with the frequency $\omega$ and assuming that the field penetrates into the planet only down to a depth $\delta$, Eq.~\eqref{dPdr} allows us to estimate the Joule heating inside it as 
\begin{equation}
P_{\rm diss} \sim \frac{2 \pi}{15} \left( \omega B_{\rm m}\right)^{2}\, \langle \sigma \rangle \, \left[ R_{\rm m}^{5} - (R_{\rm m}-\delta)^{5} \right], 
\label{induction_heating}
\end{equation}
where $R_{\rm m}$ is the radius of the planet, $\langle \sigma \rangle$ the mean value of its electric conductivity over the layer of depth $\delta$ where the external time-varying magnetic field can penetrate, and $B_{\rm m}$ is the amplitude of the time-varying magnetic field. 

The stellar magnetic field at the distance of the planet is dominated by its dipole component because it is the component with the slowest decay with the distance $r$ and $r/R$ is approximately $5-15$ in our systems considering that the putative planet can be orbiting in the middle of the quiescent radio lobes of LSR~J1835+3259. If the stellar dipole is oriented along the stellar rotation axis and the orbit of the planet is in the equatorial plane of the star, the planet will not experience any variation in the magnetic flux and no induction currents will be produced. In other words, such a configuration cannot produce any internal heating of the planet. To have induction heating, we need a dipole field inclined to the stellar spin axis and/or an orbit of the planet  inclined to the magnetic equator. For the sake of simplicity, we assume an  angle $I$ between the magnetic dipole axis of the star and the orbital angular momentum of the planet. Considering the variation in the modulus of the magnetic field along the orbit of the planet assumed to be circular, we find
\begin{equation}
    B_{\rm m} = B_{\rm s} \left( \sqrt{3 \sin^{2}I+1}-1 \right) \left( \frac{a}{R_{\rm s}}\right)^{-3},
\end{equation}
where $B_{\rm s}$ is the stellar surface dipole field at the equator, $a$ the orbit semimajor axis of the planet, and $R_{\rm s}$ the radius of the star.

\subsection{Electric conductivity of the planet}
\label{electric_conduc}
A critical parameter of our model is the mean electric conductivity of the planet  because it determines both the skin depth $\delta$ and the power dissipated by the induced currents inside the planet. Electric conductivity  is a function of the mineralogical composition of the mantle of the planet and depends on the local temperature and density. A detailed account in the case of the Earth mantle has been provided by \citet{YoshinoKatsura13}. The electric conductivity increases toward the interior as a consequence of the increase in the temperature and density. In the case of a planet  similar to Io, the internal pressure is smaller than at a similar depth inside the Earth's mantle. Considering a roughly similar density in the outer layers, the internal pressure increases proportionally to the gravity that is approximately 0.18 that of the Earth.  Therefore, we prudently postulate that the mean electric conductivity inside our putative planet is at least  $10^{-2}$ mho~m$^{-1}$, which is the conductivity of the Earth's mantle at a depth between 250 and 450~km. A more massive planet can have a higher conductivity because its internal pressure and temperature are expected to be more similar to those of the deep Earth mantle where $\sigma \sim 0.3$ mho~m$^{-1}$, at a depth of 800~km. 

The electromagnetic induction heating increases remarkably if the planet has a deep salty ocean because the electric conductivity in that case can be on the order of $\sim 3$ mho~m$^{-1}$, if we assume a salt concentration similar to that in the Earth oceans. However, due to the continuous mass loss from the surface of our planet, the lifetime of such an ocean could be shorter than the age of the system making such a scenario unlikely (cf. Sect.~\ref{discussion}). 

\section{Energetics that source radio flux}
\label{app:figure_construction}

Figure~\ref{fig:magnetosphere_sims} presents six schematic toy models that describe the necessary energetics needed to power bursty ECM emission that peaks at $\sim 2500$~$\mu$Jy \citep{Kaoetal23}, which was observed for LSR~J1835+3259 at 8.4~GHz as described in Section \ref{discussion} and Table \ref{table_comp}. In Figure~\ref{fig:magnetosphere_sims} we  also include the integrated quiescent flux at 8.4~GHz, which in very long-baseline interferometry Epoch I is reported at $F_{\nu, I} = 965 \pm 12$ $\mu$Jy and Epoch II at $F_{\nu, II} = 537 \pm 123$ $\mu$Jy \citep{Kaoetal23}. We recall that the simulation of the quiescent emission is outside the scope of the present work. Nevertheless, we added its observed geometry for completeness (see the black isocontours in Fig.~\ref{fig:magnetosphere_sims}).

The energetics of the ECM emissions are described by an ion escape parameter $\dot{M}$ (in kg~s$^{-1}$) (side pink-orange-yellow  colorbar in Fig.~\ref{fig:magnetosphere_sims}) showing either radially isotropic or toroidal. The mass loss in turn powers the total radio flux, as described in Section \ref{application}. The toy model simulates a $80$~R$_{\rm J} \times 80$~R$_{\rm J}$ field of view. 
Contours are a fixed observational reference tracing the
resolved quiescent morphology of \citet{Kaoetal23}, whose origin is not a focus of this work.  
Each panel corresponds to an illustrative plasma source scenario and is not intended for reproducing the exact auroral flux values, rather their order of magnitudes approximating the range of their mass losses for each physical process:

\begin{enumerate}
  \item {Auroral oval}
        (ionospheric--magnetospheric coupling,
        closed magnetosphere;
        Sect.~\ref{ionospheric-magnetospheric}):\\
        $\dot{M}=5\times10^{4}$\,kg\,s$^{-1}$;
        $S \sim 800\,\mu$Jy

  \item {Planet footprints}
        (Alfv\'en-wing star--planet interaction;
        Sect.~\ref{density_from_ECM}):\\
        $\dot{M}=5\times10^{4}$\,kg\,s$^{-1}$;
        $S\sim800\,\mu$Jy;
        Io-sized planet at $10\,R_{\rm J}$.

  \item {Volcanic planet}
        (tidal heating, toroidal plasma torus;
        Sect.~\ref{tidal_heating_mass_loss}):\\
        $\dot{M}=10^{5}$\,kg\,s$^{-1}$;
        $S \sim1000\,\mu$Jy;
        Io-sized planet at $10\,R_{\rm J}$.

  \item {Volcanic satellite}
        (tidally heated moon;    Sects.~\ref{planet_mass_loss_mechanisms} and~\ref{discussion}):\\
        $\dot{M}=10^{3}$\,kg\,s$^{-1}$;
        $S \sim 300\,\mu$Jy;
        planet at $10\,R_{\rm J}$ with exomoon companion.

  \item {Stellar ionospheric outflow}
        (stellar ionospheric outflow;
        Sect.~\ref{stellar_wind_main} and Appendix~\ref{stellar_wind}):\\
        $\dot{M}=10^{4}$\,kg\,s$^{-1}$;
        $S \sim 500\,\mu$Jy; no companion.

  \item {Induced electric currents}
        (electromagnetic induction heating of the planet; Sects.~\ref{planet_mass_loss_mechanisms} and~Appendix~\ref{induction_model}):\\
        $\dot{M}=4.7\times10^{3}$\,kg\,s$^{-1}$;
        $S \sim 100\,\mu$Jy;
        Io-sized planet at $10\,R_{\rm J}$.
\end{enumerate}
While the peak ECM emission, $S$, is modeled in the main text as the instantaneous burst, the image in Fig.~\ref{fig:magnetosphere_sims} seeks to approximate the beam-convolved, time-integrated $\sim$ 20 minute observation of the 8.4 GHz aurorae of the UCD.

\subsection{Beam convolution of the auroral component}
\label{app:beam}

The intrinsic auroral emission in each panel is a point
source placed at the inferred magnetic-pole positions
($|Y_R|=0.8\,R_{\star}$, $X_R\approx0$), convolved with
the Epoch~2 synthesized beam of \citet{Kaoetal23}:
\begin{equation}
  \theta_{\rm maj}=1.71~\text{mas},\quad
  \theta_{\rm min}=0.58~\text{mas},\quad
  \text{PA}=-25^{\circ}.
  \label{eq:beam_fwhm}
\end{equation}
Using the angular-to-physical scale
$1~\text{mas}=12\,R_{\rm J}$ appropriate for the distance
of LSR~J1835+3259, the full width at half maximum (FWHM) values convert to
$20.5\,R_{\rm J}$ (major) and $7.0\,R_{\rm J}$ (minor).
Converting FWHM to Gaussian standard deviations via
$\sigma=\mathrm{FWHM}/(2\sqrt{2\ln2})\approx\mathrm{FWHM}/2.355$
gives
\begin{align}
  \sigma_{\rm maj}
    &= \frac{1.71\times12}{2.355}\,R_{\rm J}
     \approx 8.7\,R_{\rm J},
  \label{eq:sigmaj}\\
  \sigma_{\rm min}
    &= \frac{0.58\times12}{2.355}\,R_{\rm J}
     \approx 2.96\,R_{\rm J}.
  \label{eq:sigmin}
\end{align}
These are passed directly to the \texttt{gaussian\_filter}
call in the toy model as pixel-equivalent sigma values
(\texttt{sigma = [sig\_b\_y * pix\_scale,
sig\_b\_x * pix\_scale]}, where
\texttt{pix\_scale = res/grid\_span = 800/80 = 10}
pixels\,$R_{\rm J}^{-1}$).
The resulting convolved integrated flux density spans $\sim16\,R_{\rm J}$,
consistent with the marginally resolved aurora detected
centrally between the two quiescent lobes in Epoch~2 of
\citet{Kaoetal23} (their Fig.~2; fitted minor axis
$\sim0.4$~mas~$\approx5\,R_{\rm J}$ within a
$2.06$~mas~$\times$~$0.55$~mas synthesized beam).

\subsection{Observational reference contours of the quiescent synchrotron emission}
\label{app:lobes}

The black contours visible in all six panels are computed
from the following two-component elliptical Gaussian
brightness map, evaluated in sky-plane coordinates
$(X,Y)$ (units of $R_{\rm J}$, positive west and north
respectively) after rotation by the magnetic-axis
inclination $\psi=-15^{\circ}$:
\begin{align}
  X_R &= X\cos\psi - Y\sin\psi,\\
  Y_R &= X\sin\psi + Y\cos\psi,
\end{align}

\begin{eqnarray}
  \mathcal{I}(X_R,Y_R)  & = &   
    A_1 \exp\!\left[ 
      -\frac{(X_R - d_0)^2}{2\sigma_{\parallel}^2} 
      - \frac{Y_R^2}{2\sigma_{\perp}^2} 
    \right] + \nonumber \\
 & &      A_2 \exp\!\left[ 
      -\frac{(X_R + d_0)^2}{2\sigma_{\parallel}^2} 
      - \frac{Y_R^2}{2\sigma_{\perp}^2} 
    \right],
  \label{eq:lobes}
\end{eqnarray}
with parameters set directly from Table~1 and Extended Data
Table~2 of \citet{Kaoetal23}:
$d_0 \approx 9.63~R_{\rm J} \approx 9\,R_{\star}$, $\sigma_{\parallel} = 4~R_{\rm J}, \
\sigma_{\perp} = 8~R_{\rm J}$, and $A_1 = 115, \quad A_2 = 100$ \quad \text{($\mu $Jy, $\, \mbox{beam}^{-1}$)}. 

The centroid offset $d_0\approx9\,R_{\star}$ is set at
measured lobe separations $17.95\pm1.39\,R_{\rm UCD}$
(Epoch~2) and $18.47\pm1.85\,R_{\rm UCD}$ (Epoch~3) from
Table~1 of \citet{Kaoetal23}.
The ratio $\sigma_{\perp}/\sigma_{\parallel}=2$ 
constrains the modeled radiation belt morphology perpendicular to the magnetic axis. Finally, the amplitude asymmetry between $A_1$ and $A_2$ \citep[representing the east and west quiescent lobes, respectively; cf. Extended Data Table~2 of][]{Kaoetal23} reproduces the observed flux contrast found in the Stokes~$I$ images of LSR~J1835+3259. We note that while the precise geometry of this schematic can be refined in future works, our current model accurately represents the energetics and integrated radio flux of each scenario.

Spatially correlated Gaussian noise with standard deviation
$\sigma_{\eta}=8$ brightness units and correlation length
$\ell_{\eta}=2$~pixels (imposed by convolution with a
circular Gaussian kernel)
is added to $\mathcal{I}$ to reproduce the clumpy morphology
of a partially resolved synchrotron source.
Contours are drawn at levels $\{30,\,60,\,90\}$ brightness
units ($\approx26\%$, $52\%$, and $78\%$ of the mean lobe
peak), and are held identical across all six panels
to emphasize that they represent a fixed observational
backdrop independent of the plasma-source scenario shown in
each panel.

For the sake of simplicity, we simulate the two lobes of the synchrotron emission as symmetric in Fig.~\ref{fig:magnetosphere_sims}. However, in the  \citet{Kaoetal23} observations (cf. their Fig.~1),  there is a clear asymmetry, with the emission from one lobe being stronger than that coming from the other, possibly as a consequence of an absorption or scattering on one side of the star where the synchrotron frequency becomes smaller or equal to the plasma frequency. 

\section{A moon orbiting a planet close to  LSR~J1835+3259}
\label{planet-moon}
We considered a planet with the mass and radius of the Earth orbiting at a distance $a_{\rm p}$ of 9.5 stellar radii from the center of LSR~J1835+3259, in the middle of its magnetosphere. If such a planet has a moon with a mass, $M_{\rm moon}$, much lower than its own mass, the semimajor axis of the orbit of the moon, $a_{\rm moon}$, must be within $0.49 \, a_{\rm H}$ if its orbit is prograde, or $0.93\,a_{\rm H}$ if it is retrograde. Here, $a_{\rm H} = a_{\rm p} [M_{\rm m}/(3M_{\rm s})]^{1/3}$ is the dynamic Hill radius of the planet, with $M_{\rm m}$ being its mass and $M_{\rm s}$ the mass of the star ($a_{\rm H}$ is not to be confused with the magnetospheric Hill radius of the star, $R_{\rm H}$, introduced in Eq.~\ref{r_hill}). These limits for orbital stability under the gravitational perturbation of the star have been derived by \citet{Domingosetal06}. We note that \citet{kisare2024} have revised the canonical 0.49 a$_H$ to 0.41 a$_H$, which only marginally affects our results below. 

The orbital period of the moon $P_{\rm moon}$ can be derived by considering the Kepler third law and that $M_{\rm moon} \ll M_{\rm m} \ll M_{\rm s}$ yielding 
\begin{equation}
P_{\rm moon} = \left( \frac{\eta^{3}}{3} \right)^{1/2} P_{\rm p},
\end{equation}
where $\eta \equiv a_{\rm moon}/a_{\rm H}$ and $P_{\rm p}$ is the orbital period of the planet around the star that is of 0.455 days in the case of our putative planet. In the case of a prograde orbiting moon, its longest orbital period is $\sim 0.2\, P_{\rm p}$ ($\eta = 0.49$), while for a retrograde orbit it is $\sim 0.53 \, P_{\rm p}$ ($\eta = 0.93$). 

The tides raised by the star on the planet will quickly circularize its orbit and synchronize its rotation with its orbital motion. Therefore, both in the case of a prograde and a retrograde orbit, the moon orbital period will be shorter than the planet rotation period. In such a regime, tides raised by the moon on the planet will produce the decay of the moon's orbit until it reaches the Roche limit of the planet where it is destroyed. The timescale for the orbital decay of the moon can be estimated using, for example, Eq.~(6) of \citet{CollierCameronJardine18}.  It can be recast as
\begin{equation}
    t_{\rm d} = \frac{Q^{\prime}_{\rm p}}{117\pi} \left[\eta^{5} \left( \frac{\eta^{3}}{3}\right)^{1/2} \right] \frac{M_{\rm m}}{M_{\rm moon}} \left( \frac{M_{\rm m}}{3M_{\rm s}} \right)^{5/3} \left(\frac{a_{\rm moon}}{R_{\rm m}} \right)^{5} P_{\rm p}, 
\end{equation}
where $Q^{\prime}_{\rm p}$ is the modified tidal quality factor of the planet, which is assumed constant along the tidal evolution of the moon orbit, and $R_{\rm m}$ is the radius of the planet. Considering a moon with a mass of $0.02$~M$_{\oplus}$, i.e., a mass comparable with that of Io, and the above parameters for the star--planet and the planet--moon systems, we find
\begin{equation}
t_{\rm d} \la 0.113 \, Q_{\rm p}^{\prime} P_{\rm p}
\end{equation}
for a prograde orbit, and 
\begin{equation}
t_{\rm d} \la 7.312 \, Q_{\rm p}^{\prime} P_{\rm p}
\end{equation}
for a retrograde orbit of the moon. Considering a moon orbiting a terrestrial  planet that has $Q^{\prime}_{\rm p} \sim 10^{3}$, the timescale for the orbital decay is on the order of a few years only. Therefore, we do not expect the moon to survive long enough to be a source for the stellar magnetosphere. 

The situation may change if the planet has a completely fluid interior because in that case the value of $Q^{\prime}_{\rm p}$ can increase by several orders of magnitude. In an optimistic approach, 
assuming an upper bound for $Q^{\prime}_{\rm p} = 10^{12}$ \citep{GoldreichNicholson77,Wu05}, we find a tidal decay timescale shorter than 0.14 and 9.1~Gyr, respectively. This is longer than the system age of 22 $\pm$ 4~Myr according to \citet{Berdyuginaetal17}, thus making the moon a potentially relevant plasma source for the system. The extremely large tidal $Q^{\prime}_{\rm p}$ by \citet{GoldreichNicholson77} is not present in our solar system, presumably due to the large energy dissipation associated with the excitation of  convective or inertial modes in the upper atmosphere or the interior of Jupiter and  Saturn. The excitation of these frequencies is likely not the case for close-in planets; therefore, \citet{Ozaetal19} reassessed the situation and found that hot Jupiters (and fluid-body planets in general) may have tidal $Q^{\prime}_{\rm p}$'s as large as 10$^{11}$, although the three-body stability problem, which also depends on eccentricity and obliquity, is complex. In \citet{Cassidyetal09} and \citet{Ozaetal19},  the circular restricted three-body problem is solved for stellar perturbations on a natural satellite, providing tidal heating from the star itself, which may significantly increase the mass loss reported here. 

The situation would become more favorable in the case of a moon orbiting a giant planet located at a greater  distance from the star. In this case, the evolution of the radius of the planet and its modified tidal quality factor could allow the moon to migrate outward and reach an orbital separation where it will no longer be affected by tides over timescales on the order of 1~Gyr \citep{Alvarado-Montesetal17,Sucerquiaetal20}. 

\end{appendix}
\end{document}